\documentclass[aip, amsmath,amssymb,uspaper,reprint]{revtex4-1} 
\usepackage{graphicx}
\usepackage{dcolumn}
\usepackage{bm}

\usepackage[utf8]{inputenc}
\usepackage[T1]{fontenc}
\usepackage{mathptmx}

\usepackage{multirow}
\usepackage{color}      
\usepackage{feynmp}
\usepackage{float}

\usepackage{ifpdf}
\ifpdf
\fi
\makeatletter
\def\endfmffile{%
  \fmfcmd{\p@rcent\space the end.^^J%
          end.^^J%
          endinput;}%
  \if@fmfio
    \immediate\closeout\@outfmf
  \fi
  \IfFileExists{\thefmffile.mp}{\immediate\write18{mpost \thefmffile}}{}
  \let\thefmffile\relax
}
\makeatother
\newcommand{\nn}{\nonumber\\}
\newcommand{\p}{\bm{p}}
\newcommand{\kk}{\bm{k}}
\newcommand{\w}{\omega}
\newcommand{\la}{\langle}
\newcommand{\ra}{\rangle}
\newcommand{\be}{\begin{equation}}
\newcommand{\ee}{\end{equation}}
\newcommand{\eqn}[1]{\label{#1}}
\newcommand{\eq}[1]{Eq.~(\ref{#1})}
\newcommand{\eqs}[1]{Eqs.~(\ref{#1})}
\newcommand{\fign}[1]{\label{#1}}
\newcommand{\fig}[1]{Fig.~\ref{#1}}
\newcommand{\figs}[1]{Figs.~\ref{#1}}
\newcommand{\f}{\bar{f}}

\newcommand{\A}{\alpha}
\newcommand{\B}{\beta}

\newcommand{\fm}{(\mbox{fm}^{-1})}

\begin{document}

\title{Dyson-Schwinger approach to pion-nucleon scattering using time-ordered perturbation theory}

\author{B.\ Blankleider}
\email{boris.blankleider@flinders.edu.au} 
\affiliation{College of Science and Engineering, Flinders University, Bedford Park, SA 5042, Australia}
\author{J.\ L.\ Wray}
\email{jordan.wray@flinders.edu.au} 
\affiliation{College of Science and Engineering, Flinders University, Bedford Park, SA 5042, Australia}
\author{A.\ N.\ Kvinikhidze}
\email{sasha\_kvinikhidze@hotmail.com}
\affiliation{A.\ Razmadze Mathematical 
Institute, Georgian Academy of Sciences, Merab Aleksidze St.\ 1, 380093 Tbilisi, Georgia}


\date{\today}

\begin{abstract}
We present a simple description of pion-nucleon ($\pi N$) scattering taking into account the full complexity of pion absorption and creation on the nucleon. To do this we solve Dyson-Schwinger equations within the framework of Time-Ordered Perturbation Theory. This enables us to construct partial wave separable $ \pi N$ t matrices that can be useful in models of nuclear processes involving fully dressed nucleons. At the same time, our approach demonstrates features of Quantum Field Theory, like particle dressing, renormalisation, and the use of Dyson-Schwinger equations, in a non-relativistic context that is maximally close to that of Quantum Mechanics. For this reason, this article may also be of pedagogical interest.
\end{abstract}

\maketitle 

\section{Introduction} 

From the late 1970's through to the early 1990's, much effort was devoted to modelling the coupled $\pi NN-NN$ system\cite{Mizutani:1977xw, Avishai:1979it,*Avishai:1979nm,*Avishai:1981dbt,Thomas:1979iw,Afnan:1980hp,*Afnan:1979tv,Blankleider:1981yp,*Afnan:1985qm,Rinat:1982ry,Mizutani:1986qt,*Lamot:1987tf,*Mizutani:1989uw,*Fayard:1992ww}. For practical reasons, almost all these works used the non-relativistic framework of Time-Ordered Perturbation Theory (TOPT) and implemented the so-called {one-pion approximation} (OPA) whereby all states containing two or more pions were neglected. The OPA allowed for the use of Faddeev-like equations for  the coupled $\pi NN-NN$ system, but contained an inconsistency in that the nucleons in $NN$ states were only partially dressed\cite{Sauer:1984um}. In this respect we note that in the OPA,  nucleons in $\pi NN$ states contain no explicit dressing at all, while the $\pi N$ input to these equations (consisting of the dressed $\pi NN$ vertex and a ``background'' $\pi N$ t matrix) allows for a fully dressed nucleon in the pole-term. This inconsistency was finally resolved by relaxing the OPA and reformulating the $\pi NN-NN$  equations in terms  convolution integrals which resulted in all nucleons in the model (including the nucleons in the $\pi N$ and $N N$ input to the model) being fully dressed\cite{Kvinikhidze:1992sv,*Kvinikhidze:1992em,Blankleider:1993dc}. Unfortunately, no numerical work on the resulting consistent formulation has been carried out so far.
Here we would like to make a modest contribution to the future solution of the convolution equations by constructing the $\pi N$  input to these equations.
This input consists of separable $\pi N$ t matrices that describe $\pi N$ scattering data, and because all nucleons are fully dressed, necessarily involves the solution of Dyson-Schwinger equations. 

In the process of carrying out this work, it became apparent that the constructed  $\pi N$ equations and their solution can also be useful from a pedagogical point of view. In particular, we have in mind that the study of Relativistic Quantum Field Theory (RQFT) forms one of the most difficult challenges in a Physics student's education. Yet some of the most fundamental concepts encountered in RQFT, like particle creation and annihilation, self-energy, renormalization, and associated equations like the Dyson-Schwinger equations, can be easily demonstrated in the non-relativistic framework of Time-Ordered Perturbation Theory (TOPT) where little more than a knowledge of standard Quantum Mechanics is needed, and in particular, where the substantial extra machinery of a relativistically covariant approach is completely avoided. The current paper is able to provide just such a demonstration on the example of a simple, yet field-theoretically complete model of pion-nucleon ($\pi N$) scattering where phenomenological separable potentials and a phenomenological bare $\pi NN$ vertex are used to fit experimental $\pi N$ phase shifts. The equations of the model are conveniently represented in diagrammatic form in \fig{fig:eq} of Sec.\ IV below, and have their origin in the works that described the coupled $NN-\pi N N$ system using few-body equations \cite{Mizutani:1977xw,Avishai:1979it,Afnan:1980hp,Blankleider:1981yp,Rinat:1982ry}; however, all these works involved the OPA where all states of two or more pions are neglected. Our innovation, therefore, is to implement a complete dressing of all nucleon propagators appearing in the model, a task that involves the numerical solution of the Dyson-Schwinger equations (as represented by \figs{fig:eq} (b), (c), and (d)), and  which, in the process, demonstrates some of the main features of RQFT presented in a simple non-relativistic setting. To enhance the pedagogical aspects of this paper, we endeavour to give a more detailed  presentation of our work than may normally be expected.

\section{Pion-nucleon model}

The pion-nucleon model we consider involves only pion and nucleon degrees of freedom. To take into account that pions can get created and absorbed by the nucleon, we shall use a Hilbert space ${\cal H}$ that is a direct sum of subspaces, each consisting of states with a different number of pions, thus
\be
{\cal H} =  {\cal H}_N\oplus {\cal H}_{\pi N}\oplus {\cal H}_{\pi \pi N}\oplus \ldots
\ee
Each of these subspaces contain free particle states which shall be labelled by the momenta of the particles (to save on notation we suppress spin and isospin labels), thus $|\p\ra \in {\cal H}_N$, 
$|\kk\, \p\ra \in {\cal H}_{\pi N}$, $|\kk_1\, \kk_2\, \p\ra \in {\cal H}_{\pi\pi N}$, etc., where $\p$ is the momentum of the nucleon and $\kk, \kk_1, \kk_2,\ldots$ are pion momenta.
 This multi-pion plus one-nucleon system is then described within a second quantization approach\cite{Fetter} where the Hamiltonian $H$ is expressed in terms of annihilation and creation operators, $a_\pi, a^\dagger_\pi$ for pions, and  $a_N, a^\dagger_N$ for nucleons, defined such that
\begin{align}
|\kk\ra = a_\pi^\dagger(\kk) | 0 \ra , \hspace{1cm} |\p\ra = a_N^\dagger(\p)| 0 \ra,
\end{align}
where $|0\ra$ is the vacuum state, and
satisfying the commutation and anti-commutation relations
\begin{subequations}  \eqn{comm}
\begin{align}
\left[a_{\pi }(\kk),a_{\pi }^{\dagger }(\kk')\right]& =\delta (\kk-\kk'), \hspace{3mm}
\left[a_{\pi }(\kk),a_{\pi }(\kk')\right]=0, \\
\left\{a_N(\p),a_N^{\dagger }(\p')\right\}& =\delta (\p-\p'), \hspace{2mm}
\left\{a_N(\p),a_N(\p')\right\}=0.
\end{align}
\end{subequations}
In particular, the free Hamiltonians for pions and nucleons are given by
\begin{subequations}  \eqn{H0}
\begin{align}
H_0^\pi &=\int \w_\pi(k) \,a_\pi^\dagger(\kk) a_\pi(\kk) \,d\kk  \eqn{H0pi}   \\
H_0^N &=\int (E_N(p) +m_0) \,a_N^\dagger(\p) a_N(\p) \,d\p   \eqn{H0N}
\end{align}
\end{subequations}
where
\begin{equation}
\w_\pi(k)=\left(\kk^2+m_{\pi }{}^2\right){}^{1/2},\hspace{1cm} E_N(p)=\frac{\p^2}{2m},    \eqn{Es}
\end{equation}
with $m_\pi$ being the mass of the pion and $m$ being the mass of the nucleon. Note that for simplicity of presentation we have used non-relativistic kinematics for the nucleon and relativistic kinematics for the pion (necessitated by the relatively small mass of the pion). Such a ``semi-relativistic'' choice of kinematics should be sufficiently accurate to describe $\pi N$ scattering for centre of mass (c.m.) energies below  $m+3m_\pi$, being approximately the experimental onset of inelasticity, and will provide a convenient connection to related works using such kinematics \cite{Thomas:1976px,McLeod:1984cu}.

Because the interactions that we shall use contribute to the physical mass $m$ of a nucleon, the rest mass used for the nucleon in \eq{H0N}, the so-called {\em bare} mass $m_0$, needs to be chosen appropriately (by contrast, our interactions do not contribute to the physical mass of the pion). 
We note that the free particle states are eigenstates of the total free Hamiltonian $H_0$,
\be
H_0 = H^\pi_0+H^N_0,
\ee
so that
\begin{align}
H_0&|\kk\ra = \w_\pi(k)|\kk\ra  , \quad\quad H_0|\p\ra = (E_N(p)+m_0)|\p\ra ,  \nn
H_0&|\kk_1\kk_2\ldots,\p\ra = [\w_\pi(k_1)+ \w_\pi(k_2)+\ldots+E_N(p)+m_0]|\p\ra .
\end{align}
The full Hamiltonian is then specified as
\be
H = H_0+H_I \eqn{H}
\ee
where $H_I$ is the interaction Hamiltonian. In this work we shall assume a model interaction Hamiltonian of the form
\begin{equation}
H_I =\int J_N(\kk) \,a_\pi^\dagger(\kk) \,d\kk + \int J_N^\dagger (\kk) \,a_\pi(\kk) \,d\kk     \eqn{HI}
\end{equation}
where
\begin{subequations}\eqn{Js}
\begin{align}
J_N(\kk) &=\int \delta(\p+\kk-\p') \frac{1}{\sqrt{w_\pi}} F_0(\p,\p') a_N^\dagger(\p) a_N(\p') \,d\p \,d\p'   \eqn{J} \\
J^\dagger_N(\kk) &=\int \delta(\p+\kk-\p') \frac{1}{\sqrt{w_\pi}} \bar{F}_0(\p',\p) a_N^\dagger(\p') a_N(\p) \,d\p \,d\p'.  \eqn{Jd}
\end{align}
\end{subequations}
In the above (and below) $\w_\pi$ is to be understood as meaning $\w_\pi(k)$.
This interaction Hamiltonian describes the elementary processes $\pi N\leftrightarrow N$ whose amplitude is given by the real function $F_0(\p,\p')=\bar{F}_0(\p',\p)$, and which we shall refer to as the {\em bare} $\pi N N$ vertex. It will be assumed that $F_0(\p,\p')$ is invariant under rotations and under space inversion, so that angular momentum and parity are conserved in the resulting description of $\pi N$ scattering.

With the model thus defined, we can now define the Green function $G$ that shall be used to describe pion-nucleon scattering, as
\begin{equation}
\delta \left(\kk +\p -\kk'-\p'\right) G(\p',\kk,\p,E) = \langle \kk'\, \p' |\frac{1}{E^+-H}|\kk\, \p\rangle.   \eqn{G}
\end{equation}
In a similar way we define the single nucleon Green function $g$, also called the {\em dressed nucleon propagator}, as 
\be
\delta (\p-\p')g(E,\p) = \langle \p'|\frac{1}{E^+-H}|\p\rangle  , \eqn{g}
\ee
the {\em bare nucleon propagtor}  $g_0$ as
\be
\delta (\p-\p')g_0(E,\p) = \langle \p'|\frac{1}{E^+-H_0}|\p\rangle   \eqn{g0}
\ee
and the Green function for the process $N\rightarrow\pi N$, $\tilde{f}$, as
\begin{equation}
\delta (\p'+\kk'-\p)\tilde{f}(\p',\p,E) =\langle \kk'\, \p'|\frac{1}{E^+-H}|\p\rangle.   \eqn{tf}
\end{equation}

To help evaluate these Green functions, it is useful to note the commutation relations
\begin{subequations}
\begin{align}
\left[a_{\pi }(\kk),H_0 \right] &= w_\pi(k) a_{\pi }(\kk), \\
\left[a_{\pi }(\kk),H_I\right]&=J_N(\kk),
\end{align}
\end{subequations}
which follow from \eqs{comm}, (\ref{H0}) and (\ref{HI}). In a similar way it follows that
\begin{subequations}
\begin{align}
J_N(\kk)|\p'\rangle 
&=\frac{1}{\sqrt{w_\pi}}F_0(\p,\p')|\p \rangle , \\
\la \p' | J^\dagger_N(\kk) &=\langle \p |\frac{1}{\sqrt{w_\pi}}\bar{F}_0(\p',\p),
\end{align}
\end{subequations}
where $\p + \kk = \p'$. Thus
\begin{subequations} \eqn{HIp}
\begin{align}
H_I |\p'\ra &= \int \delta(\p+\kk-\p')\frac{1}{\sqrt{\omega _\pi}}F_0(\p ,\p')|\kk\, \p \rangle\, d\kk\, d\p \\
\la \p'| H_I  &= \int \delta(\p+\kk-\p')\la \kk\, \p | \frac{1}{\sqrt{\omega _\pi}}\bar{F}_0(\p' ,\p)\, d\kk\, d\p. 
\end{align}
\end{subequations}

\section{Time-ordered perturbation theory}

Despite the relative simplicity of the quantum field theoretic pion-nucleon model introduced above, there is no known way to calculate any of the Green functions of \eqs{G}, (\ref{g}), or (\ref{tf}) exactly. We therefore follow the usual path taken in both quantum mechanics and in quantum field theory, and introduce a perturbation expansion. In our case this means expanding the Green function operator $1/(E^+-H)$ into a series around the free Green function operator $1/(E^+-H_0)$: 
\begin{align}
\frac{1}{E^+-H}&=\frac{1}{E^+-H_0}+\frac{1}{E^+-H_0}H_I\frac{1}{E^+-H_0}\nn
& +\frac{1}{E^+-H_0}H_I\frac{1}{E^+-H_0}H_I\frac{1}{E^+-H_0}+\ldots   \eqn{pert}
\end{align}
This expansion defines \textit{ time-ordered perturbation theory }(TOPT), and has the feature that momentum matrix elements of each term of the expansion can be represented
graphically by a ``perturbation diagram''. For example, if one evaluates this expansion for the matrix element defining the dressed nucleon propagator $g(E)$ of \eq{g}, one obtains a series whose graphical representation is given in \fig{fig:g}. The diagrammatic rules that relate these diagrams to mathematical expressions can be found by evaluating the momentum matrix elements of just the first three terms of \eq{pert}.  With the help of \eqs{HIp} one obtains
\begin{widetext}
\begin{align}
&\delta(\p-\p')g(E,\p) =\la  \p' |\frac{1}{E^+-H_0}|\p \ra +\la \p' |\frac{1}{E^+-H_0}H_I\frac{1}{E^+-H_0}|\p \ra
+\la  \p' |\frac{1}{E^+-H_0}H_I\frac{1}{E^+-H_0}H_I\frac{1}{E^+-H_0}|\p \ra + \ldots \nn[3mm]
&=\delta (\p -\p ')g_0\left(E-E_p\right) + 0 + \delta(\p -\p')g_0(E,\p')
\int  \frac{ \frac{1}{\sqrt{\omega _\pi}}\bar{F}_0(\p' ,\p'') \frac{1}{\sqrt{\w _\pi}}F_0(\p'',\p) }{E^+-E_N(p'')-m_0-\w_\pi(k) } d\kk\,\,
 g_0(E,\p)\, +\ldots
\end{align}
\end{widetext}
where $\p'' = \p -\kk$, so that
\begin{align}
g&(E,\p) = g_0(E,\p)
+ g_0(E,\p) \int \bar{f}_0(\p,\kk \p'') \nn
&\times  \frac{1}{E^+-E_N(p'')-m_0-\w_\pi(k) } 
f_0(\kk \p'',\p) g_0(E,\p)\, d\kk  + \dots   \eqn{gseries}
\end{align}
\begin{figure}[t]
\begin{center}
\begin{fmffile}{Gammag}
\fmfset{dash_len}{2.8mm}
\begin{align*}
&
\parbox{13mm}{
\begin{fmfgraph*}(13,14)
\fmfleft{i} \fmfright{f} 
\fmf{plain}{i,m,f}
\fmfv{decor.shape=circle,decor.filled=full,decor.size=1.4mm}{m}
\fmfv{label=$g$,l.a=-90}{m}
\end{fmfgraph*}} \  \ =\  \
\parbox{11mm}{
\begin{fmfgraph*}(11,14)
\fmfleft{i} \fmfright{f}
\fmf{plain}{i,m,f}
\fmfv{label=$g_0$,l.a=-90}{m}
\end{fmfgraph*}} \ \ +\  \
\parbox{18mm}{
\begin{fmfgraph*}(17,14)
\fmfleft{i} \fmfright{f}
\fmf{plain}{i,v1,m,v2,f}
\fmf{dashes,left=.9,tension=0}{v1,v2}
\fmfv{decor.shape=circle,decor.filled=empty,decor.size=1.4mm}{v1}
\fmfv{decor.shape=circle,decor.filled=empty,decor.size=1.4mm}{v2}
\fmfv{label=$\f_0$,l.a=-90}{v1}
\fmfv{label=$f_0$,l.a=-90}{v2}
\end{fmfgraph*}}
\  \ +\  \
\parbox{22mm}{
\begin{fmfgraph*}(22,14)
\fmfleft{i} \fmfright{f}
\fmf{plain}{i,v1,m1,m2,v2,f}
\fmf{dashes,left=.9,tension=0}{v1,m2}
\fmf{dashes,right=.9,tension=0}{v2,m1}
\fmfv{decor.shape=circle,decor.filled=empty,decor.size=1.4mm}{v1}
\fmfv{decor.shape=circle,decor.filled=empty,decor.size=1.4mm}{v2}
\fmfv{decor.shape=circle,decor.filled=empty,decor.size=1.4mm}{m1}
\fmfv{decor.shape=circle,decor.filled=empty,decor.size=1.4mm}{m2}
\end{fmfgraph*}} \nn
&\  \ +\  \
\parbox{24mm}{
\begin{fmfgraph*}(24,14)
\fmfleft{i} \fmfright{f}
\fmf{plain}{i,v1,m1,m2,v2,f}
\fmf{dashes,left=.75,tension=0}{v1,v2}
\fmf{dashes,left=.97,tension=0}{m1,m2}
\fmfv{decor.shape=circle,decor.filled=empty,decor.size=1.4mm}{v1}
\fmfv{decor.shape=circle,decor.filled=empty,decor.size=1.4mm}{v2}
\fmfv{decor.shape=circle,decor.filled=empty,decor.size=1.4mm}{m1}
\fmfv{decor.shape=circle,decor.filled=empty,decor.size=1.4mm}{m2}
\end{fmfgraph*}}
\  \ +\  \
\parbox{35mm}{
\begin{fmfgraph*}(35,14)
\fmfleft{i} \fmfright{f} 
\fmf{plain}{i,v1,m1,m2,v2,f}
\fmf{dashes,left,tension=0}{v1,m1}
\fmf{dashes,left,tension=0}{m2,v2}
\fmfv{decor.shape=circle,decor.filled=empty,decor.size=1.4mm}{m1}
\fmfv{decor.shape=circle,decor.filled=empty,decor.size=1.4mm}{m2}
\fmfv{decor.shape=circle,decor.filled=empty,decor.size=1.4mm}{v1}
\fmfv{decor.shape=circle,decor.filled=empty,decor.size=1.4mm}{v2}
\end{fmfgraph*}}
\  \ +\  \ \dots
\end{align*}
\end{fmffile}
\end{center}
\vspace{-1cm}
\caption{\fign{fig:g} Graphical representation of  the perturbation expansion of the dressed nucleon propagator $g$. The corresponding mathematical expression is given in \eq{gseries}.}
\end{figure}
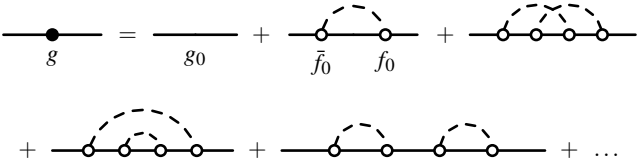
where
\begin{subequations}\eqn{f0}
\begin{align}
f_0(\kk \p'',\p) &= \frac{1}{\sqrt{\w_\pi}}F_0(\p'',\p), \\ 
\f_0(\kk \p,\p'') &= \frac{1}{\sqrt{\w _\pi}}\bar{F}_0(\p,\p''). 
\end{align}
\end{subequations}
Comparing \fig{fig:g} and \eq{gseries}, it is evident that a solitary solid line corresponds to the bare nucleon propagator $g_0$, every open-circle $N\rightarrow \pi N$ $(\pi N\rightarrow N)$ vertex in \fig{fig:g} corresponds to the bare vertex function $f_0$ $(\bar{f}_0)$ defined in \eqs{f0}, every intermediate state of solid (nucleon) and dashed (pion) lines corresponds to a propagator whose mathematical expression is given by the appropriate matrix elements of $1/(E^+-H_0)$, and every closed loop corresponds to an integral over the pion momentum.

\section{pion-nucleon equations}

As the Green function $G(\p',\kk,\p,E) $ describing $\pi N$ scattering can be equated to a complete sum of perturbation diagrams, this provides the opportunity to rearrange this sum so as to express it completely in terms of useful quantities like potentials, t matrices, and other Green functions, thereby generating scattering equations of a similar nature to those found in Quantum Mechanics (e.g. the Lippmann-Schwinger equation).  Such a rearrangement leads to a set of coupled equations for $\pi N$ scattering illustrated in \fig{fig:eq}. First derived by Mizutani and Koltun  using Feshbach projection operators \cite{Mizutani:1977xw}, these equations were later derived in the same context of TOPT as used here \cite{Afnan:1980hp} and also in the context of RQFT \cite{Mizutani:1981cb}. Here we shall give a brief derivation following the arguments used in Ref.\  \cite{Afnan:1980hp}.

We start by writing the Green function  $G(\p',\kk,\p,E) $ in operator form as
\be
G(E) = G_0(E) + G_0(E) t(E) G_0(E)              \eqn{Gt}
\ee
where $G_0(E)$ is the ``dressed $\pi N$ propagator'' consisting of all the disconnected diagrams of $G(E)$, and $t(E)$ is the $\pi N\rightarrow \pi N$ t matrix which is defined by this equation. Note that each quantity in \eq{Gt} is an operator acting in subspace ${\cal H}_{\pi N}$, with $G(E)$ being specifically defined such that
\begin{align}
 \langle \kk'\, \p' |G(E)|\kk\, \p\rangle &= \langle \kk'\, \p' |\frac{1}{E^+-H}|\kk\, \p\rangle.   \eqn{hG}
\end{align}
It is evident that the term $G_0(E)t(E)G_0(E)$ in \eq{Gt} consists of all possible {\em connected}  $\pi N\rightarrow \pi N$ diagrams and that $t(E)$ consists of the same diagrams but with no attached initial- and final-state  $\pi N$ propagators, colloquially referred to as diagrams with ``chopped legs''.

Further progress can be made by defining a ``background'' $\pi N$ t matrix  $t^{(1)}(E)$ as the sum of all diagrams of $t(E)$ that have one or more pions in every intermediate state, as one can then write
\be
t(E) = f(E) g(E) \f(E) + t^{(1)}(E)   \eqn{t}
\ee
where $f(E)$  $(\f(E))$ is the ``dressed vertex'' consisting of all possible  $N\rightarrow \pi N$ ($\pi N\rightarrow N$) chopped-leg diagrams with at least one pion in every intermediate state. Similarly, one can define  t matrix  $t^{(2)}(E)$ as the sum of all diagrams of $t(E)$ that have two or more pions in every intermediate state, in which case we can write Lippmann-Schwinger-like equations
\begin{subequations} \eqn{t1}
\begin{align}
 t^{(1)}(E) &= v(E) +  v(E) G_0(E)  t^{(1)}(E), \eqn{tb} \\
 &= v(E) +  t^{(1)} G_0(E)  v(E)
 \end{align}
 \end{subequations}
where $v(E) \equiv t^{(2)}(E)$ plays the role of a $\pi N \rightarrow \pi N$ potential. Using a similar argument, one can obtain the equations
\begin{subequations}\eqn{fs}
\begin{align}
f(E) &= f_0(E) + t^{(1)}(E) G_0(E) f_0(E), \eqn{f} \\
\f(E) &= \f_0(E) + \f_0(E) G_0(E) t^{(1)}(E), 
\end{align}
\end{subequations}
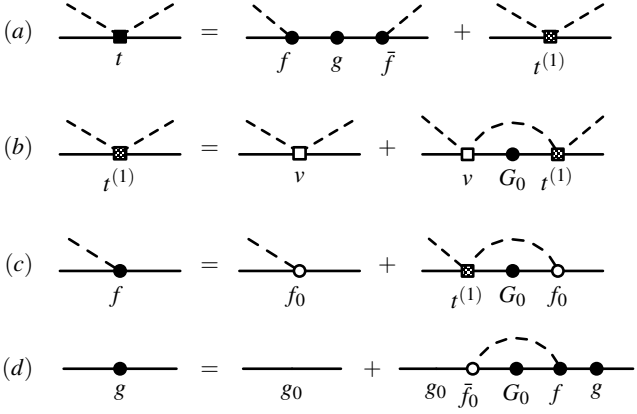
\begin{figure}[t]
\begin{fmffile}{t_piN}
\begin{align}
(a)\hspace*{-2mm} &  &
\begin{fmfgraph*}(20,5)
\fmfleft{fN,fpi}\fmfright{iN,ipi}
\fmf{plain}{iN,vN,fN}
\fmffreeze
\fmf{dashes,tension=.15}{vpi,fpi}
\fmf{dashes,tension=.15}{vpi,ipi}
\fmf{phantom,tension=10}{vpi,v,vN}
\fmfv{label=$t$,l.d=.4h,decor.shape=square,decor.filled=full,decor.size=1.5mm}{v}
\end{fmfgraph*}
& =
\begin{fmfgraph*}(30,5)
\fmfleft{fn,fpi}\fmfright{in,ipi}
\fmf{plain}{in,n2}
\fmf{plain}{n2,n,n1}
\fmf{plain}{n1,fn}
\fmffreeze
\fmf{dashes}{n1,fpi}
\fmf{dashes}{n2,ipi}
\fmfv{label=$f\,,\,, $,l.a=-112,l.d=.4h,decor.shape=circle,decor.filled=full,decor.size=1.5mm}{n1}
\fmfv{label=$g$,l.d=.5h,decor.shape=circle,decor.filled=full,decor.size=1.5mm}{n}
\fmfv{label=$\f$,l.a=-68,l.d=.4h,decor.shape=circle,decor.filled=full,decor.size=1.5mm}{n2}
\end{fmfgraph*}
+
\begin{fmfgraph*}(20,5)
\fmfleft{fN,fpi}\fmfright{iN,ipi}
\fmf{plain}{iN,vN,fN}
\fmffreeze
\fmf{dashes,tension=.15}{vpi,fpi}
\fmf{dashes,tension=.15}{vpi,ipi}
\fmf{phantom,tension=10}{vpi,v,vN}
\fmfv{label=$t^{(1)}$,l.d=.4h,decor.shape=square,decor.filled=gray50,decor.size=1.5mm}{v}
\end{fmfgraph*}
\nn[8mm]
(b)\hspace*{-2mm} & &
\begin{fmfgraph*}(20,5)
\fmfleft{fN,fpi}\fmfright{iN,ipi}
\fmf{plain}{iN,vN,fN}
\fmffreeze
\fmf{dashes,tension=.15}{vpi,fpi}
\fmf{dashes,tension=.15}{vpi,ipi}
\fmf{phantom,tension=10}{vpi,v,vN}
\fmfv{label=$t^{(1)}$,l.d=.4h,decor.shape=square,decor.filled=gray50,decor.size=1.6mm}{v}
\end{fmfgraph*}
& =
\begin{fmfgraph*}(20,5)
\fmfleft{fN,fpi}\fmfright{iN,ipi}
\fmf{plain}{iN,vN,fN}
\fmffreeze
\fmf{dashes,tension=.15}{vpi,fpi}
\fmf{dashes,tension=.15}{vpi,ipi}
\fmf{phantom,tension=10}{vpi,v,vN}
\fmfv{label=$v$,l.d=.5h,decor.shape=square,decor.filled=empty,decor.size=1.6mm}{v}
\end{fmfgraph*}
+
\begin{fmfgraph*}(30,5)
\fmfleft{fn,fpi}\fmfright{in,ipi}
\fmf{plain}{in,n1}
\fmf{plain,tension=1}{n1,n,n2}
\fmf{plain}{n2,fn}
\fmffreeze
\fmf{dashes}{ipi,n1}
\fmf{dashes,right=.7}{n1,n2}
\fmf{dashes}{fpi,n2}
\fmfv{label=$v$,l.a=-90,l.d=.55h,decor.shape=square,decor.filled=empty,decor.size=1.5mm}{n2}
\fmfv{label=$G_0$,l.a=-90,l.d=.4h,decor.shape=circle,decor.filled=full,decor.size=1.5mm}{n}
\fmfv{label=$t^{(1)}$,l.a=-90,l.d=.35h,decor.shape=square,decor.filled=gray50,decor.size=1.5mm}{n1}
\end{fmfgraph*}
\nn[8mm]
(c)\hspace*{-2mm} & &
\begin{fmfgraph*}(20,5)
\fmfleft{fn,fpi}\fmfright{in,ipi}
\fmf{plain}{in,n}
\fmf{plain}{n,fn}
\fmffreeze
\fmf{dashes}{n,fpi}
\fmfv{label=$f\,,\,, $,l.a=-95,l.d=.4h,decor.shape=circle,decor.filled=full,decor.size=1.5mm}{n}
\end{fmfgraph*}
& =
\begin{fmfgraph*}(20,5)
\fmfleft{fn,fpi}\fmfright{in,ipi}
\fmf{plain}{in,n}
\fmf{plain}{n,fn}
\fmffreeze
\fmf{dashes}{n,fpi}
\fmfv{label=$f_0\,,\,, $,l.a=-95,l.d=.4h,decor.shape=circle,decor.filled=empty,decor.size=1.5mm}{n}
\end{fmfgraph*}
+
\begin{fmfgraph*}(30,5)
\fmfleft{fn,fpi}\fmfright{in,ipi}
\fmf{plain}{in,n1}
\fmf{plain,tension=1}{n1,n,n2}
\fmf{plain}{n2,fn}
\fmffreeze
\fmf{dashes,right=.7}{n1,n2}
\fmf{dashes}{n2,fpi}
\fmfv{decor.shape=circle,decor.filled=empty,decor.size=1.5mm}{n1}
\fmfv{label=$t^{(1)}$,l.a=-90,l.d=.4h,decor.shape=square,decor.filled=gray50,decor.size=1.5mm}{n2}
\fmfv{label=$G_0$,l.a=-90,l.d=.4h,decor.shape=circle,decor.filled=full,decor.size=1.5mm}{n}
\fmfv{label=$f_0$,l.a=-90,l.d=.4h,decor.shape=circle,decor.filled=empty,decor.size=1.5mm}{n1}
\end{fmfgraph*}
\nn[8mm]
(d)\hspace*{-2mm} & &
\parbox{20mm}{
\begin{fmfgraph*}(15,5)
\fmfleft{i} \fmfright{f} 
\fmf{plain}{i,m,f}
\fmfv{decor.shape=circle,decor.filled=full,decor.size=1.5mm}{m}
\fmfv{label=$g$,l.a=-90}{m}
\end{fmfgraph*}}  &=
\parbox{18mm}{
\begin{fmfgraph*}(13,5)
\fmfleft{i} \fmfright{f}
\fmf{plain}{i,m,f}
\fmfv{label=$g_0$,l.a=-90}{m}
\end{fmfgraph*}}  + \hspace{2mm}
\parbox{31mm}{
\begin{fmfgraph*}(31,5)
\fmfleft{fn}\fmfright{in}
\fmf{plain,tension=1.2}{in,n3,n1}
\fmf{plain, tension=1}{n1,n,n2}
\fmf{plain,tension=1.2}{n2,n4,fn}
\fmffreeze
\fmf{dashes,right=.7}{n1,n2}
\fmfv{label=$f\,,\,, $,l.a=-95,l.d=.4h,decor.shape=circle,decor.filled=full,decor.size=1.5mm}{n1}
\fmfv{label=$\f_0\,,\,, $,l.a=-95,l.d=.4h,decor.shape=circle,decor.filled=empty,decor.size=1.5mm}{n2}
\fmfv{label=$G_0$,l.a=-90,l.d=.4h,decor.shape=circle,decor.filled=full,decor.size=1.5mm}{n}
\fmfv{label=$g$,l.a=-90,l.d=.4h,decor.shape=circle,decor.filled=full,decor.size=1.5mm}{n3}
\fmfv{label=$g_0$,l.a=-90,l.d=.4h}{n4}
\end{fmfgraph*} }\nonumber
\end{align}
\end{fmffile}
\caption{\fign{fig:eq} Illustration of the $\pi N$ scattering equations: (a)  The $\pi N$ t matrix expressed by \eq{t}, (b) the ``background'' $\pi N$ t matrix as given by \eq{tb}, (c) the dressed $\pi N N$  vertex   as given by \eq{f}, and (d) the dressed nucleon propagator as given by \eq{gg} whose self-energy term $\Sigma$ is expressed as in \eq{siga}.}
\end{figure}
where $f_0(E) \equiv f^{(2)}(E)$  ($\f_0(E) \equiv \f^{(2)}(E)$) is the ``bare vertex'' consisting of all possible $N \rightarrow \pi N$ ($\pi N \rightarrow N$ ) chopped-leg diagrams with at least two pions in every intermediate state. Finally, one can similarly write
\be
g(E) = g_0(E) + g_0(E) \Sigma(E) g(E)   \eqn{gg}
\ee
where
\begin{subequations}  \eqn{sig}
\begin{align}
\Sigma(E) & = \f_0(E) G_0(E) f(E), \eqn{siga}\\
& = \f(E) G_0(E) f_0(E),
\end{align}
\end{subequations}
is the nucleon ``self-energy'' or ``dressing'' term consisting of all diagrams of $g(E)$ with at least one pion in every intermediate state, but with chopped legs.  The set of equations consisting of \eqs{t}-(\ref{sig}) are illustrated in \fig{fig:eq}, and provide an exact and useful way of expressing the $\pi N$ t matrix $t(E)$. 

 In the context of RQFT, \eq{gg} (illustrated in \fig{fig:eq}(d)) is known as the Dyson equation, while the coupled set of equations \eq{tb}, \eq{f}, and \eq{gg}, illustrated in \fig{fig:eq}(b)-(d), are known as the Dyson-Schwinger (DS) equations,    
 and that is how we shall refer to the TOPT versions of these equations here. A feature of the Dyson equation is the fact that the dressed nucleon propagator $g$ is expressed in terms of the self-energy term $\Sigma$ which itself is expressed in terms of $g$  via the $
\pi N$ propagator $G_0(E)$. Such self-referencing also occurs for the background t matrix $t^{(1)}$ and the dressed vertex $f$ in the DS equations, a feature that makes these equations embody a lot of physics even in the case where the bare vertex $f_0$ and background potential $v$ are modelled phenomenologically, as will be the case in the next section.

\section{Solving the Dyson-Schwinger equations}

Here we shall follow an often used procedure where the bare $\pi N N$ vertex $f_0$ and the ``background'' $\pi N$ potential $v$ are modelled by energy-independent parametrized phenomenological functions; however, unlike in all such previous models\cite{Mizutani:1977xw,Avishai:1979it,Afnan:1980hp,Blankleider:1981yp,Rinat:1982ry,McLeod:1984cu}, we shall not be using the approximation where the exact $\pi N$ propagator $G_0(E,\kk,\p)$, defined as
\be
\delta(\kk'-\kk)\delta(\p'-\p) G_0(E,\kk,\p) = \langle \kk'\, \p' |\frac{1}{E^+-H}|\kk\, \p\rangle_{disc},
 \ee
 is modelled by the pole term $1/[E^+-E_N(p)-m-\w_\pi(k)]$; rather, we shall retain its full exact form, which in the model specified by the Hamiltonian of \eqs{H0} and \eq{HI}, is given by
\be
G_0(E,\kk,\p) = g[E-E_N(p)-\w_\pi(k)].    \eqn{G01}
\ee
As mentioned above, it is this exact form for $G_0$ which gives the DS equations the property of retaining a rich amount of physics despite what may be lost by taking phenomenological forms for $f_0$ and $v$. 

\subsection{Partial wave equations}

It is convenient to solve the Dyson-Schwinger set of equations, given in operator form in \eqs{t1} - (\ref{sig}), in the centre of mass (c.m.) of the $\pi N$ system, so that $\kk +\p = \kk'+\p'={\bf 0}$, in which case \eq{G01} can be expressed as
\be
G_0(E,\kk) = g(E-\w_k)    \eqn{G02}
\ee
where
\be
\w_k=E_N(k)+\w_\pi(k) .  \eqn{wk}
\ee
In order to reduce the dimension of these equations from 3 to 1, we shall work in the partial wave representation using the basis states
\begin{align}
| l  s j t, m_j m_t, &k \ra = \sum_{\substack{m_{t_1}  m_{t_2}\\ m_l\, m_s}} (l  m_l s  m_s|  j m_j ) (t_1 m_{t_1}t_2 m_{t_2}| t m_t )\nn
&\times \int d \hat{k}\, Y_{l m_l}(\hat{k})\, | t_1 m_{t_1}t_2 m_{t_2},\kk \ra \eqn{pw}
\end{align}
where $t_1=1$ is the isospin of the pion, $t_2=1/2$ is the isospin of the nucleon, $s=1/2$ is the spin of the nucleon,  $l$ specifies the $\pi N$ relative orbital angular momentum, $t$ is the total isospin,  and $j$ is the total angular momentum. By construction, the model Hamiltonian of \eq{H} is invariant under rotations, which implies that all matrix elements using the above partial wave basis states will not depend on the magnetic quantum numbers $m_j$ and $m_t$. Similarly, the model Hamiltonian is chosen to be invariant under space inversion, thus ensuring parity is conserved in our model which in turn, leads to $\pi N$ partial wave matrix elements that preserve the value of $l$. We shall refer to the partial wave specified by the quantum numbers $\{l j t\}$ using the usual (for $\pi N$ scattering) spectroscopic notation of the form ``$L_{(2t)(2j)}$''. Because the nucleon has quantum numbers $t=j=1/2$ and the pion has intrinsic parity of $-1$, it follows that the first term on the right hand side (RHS) of 
\eq{t}, the so-called {\em nucleon pole term}, contributes only in the $P_{11}$ partial wave. Likewise, the background $\pi N$ t matrix $t^{(1)}$ appearing in the expression for the dressed $\pi N N$ vertices of \eqs{fs} is the one in the $P_{11}$ partial wave. Thus, restricting the discussion to $\pi N$ scattering in the  $P_{11}$ partial wave, we can write the numerical form of the partial wave equations  corresponding to \fig{fig:eq} as
\begin{subequations} \eqn{dspw}
\begin{align}
t(k',k,E)&=f(k,E) g(E) \f(k,E) + t^{(1)}(k',k,E) \eqn{tpw} \\[2mm]
 t^{(1)}(k',k,E) &=v(k',k) + \int_0^\infty dq\, q^2\, v(k',q) g(E-\w_q)  t^{(1)}(q,k,E) \eqn{lse}\\[2mm]
 f(k,E) &= f_0(k) +  \int_0^\infty dq\, q^2\, t^{(1)}(k,q,E) g(E-\w_q) f_0(q) \eqn{fpw}\\[2mm]
 g(E) &= \frac{1}{E^+ -m_0-\Sigma(E)} \eqn{gexp} \\[2mm]
 \Sigma(E) &= \int_0^\infty dq\, q^2\, f_0(q) g(E-\w_q) f(q,E)
\end{align}
\end{subequations}
where partial wave labels have been omitted to save on notation. For scattering in partial waves other than $P_{11}$, essentially the same equations would apply, the only differences being that the pole term in \eq{tpw} would not appear, and one would need to distinguish the $t^{(1)}$ appearing in \eq{tpw} from the  $P_{11}$ partial wave  $t^{(1)}$ appearing in \eq{fpw} .

In order to help solve these equations numerically, it is useful to know the analytic structure of the dressed nucleon propagator $g(E)$. 
As a basic requirement of the theory, $g(E)$ must  have a pole at the physical nucleon mass $m$. To ensure this, the bare mass is set to the value $m_0=m-\Sigma(m)$ 
so that
\be
g(E) = \frac{1}{E^+-m-\Sigma(E)+\Sigma(m)}
\ee
and therefore
\be
g(E) \mathop{\sim}_{E\rightarrow m}\ \frac{Z}{E^+-m}
\ee
where $Z$ is the wave function renormalization constant given by
\be
Z=\frac{1}{1-\Sigma'(m)}
\ee
where the prime indicates a derivative with respect to $E$. To evaluate $\Sigma'(m)$, one can use the identity
\be
 \Sigma'(E) = \int_0^\infty dq\, q^2\, \f(q,E) g'(E-\w_q) f(q,E).  \eqn{Sigmap}
\ee
which can be easily proved using the operator form of \eqs{dspw}.

Besides a nucleon pole, it can be shown that $g(E)$ also contains a cut starting at $E=m+m_\pi$ and extending to $+\infty$, and that this analytic structure implies the following ``dispersion relation'' \cite{Kvinikhidze:1992em}
\be
g(E)=\frac{Z}{E^+-m} - \frac{1}{\pi} \int_{m+m_\pi}^\infty \frac{\mbox{Im} g(\omega)}{E^+-\w} d\w.  \eqn{disp}
\ee
As we shall see, this relationship between $g(E)$ and its imaginary part will prove very useful in the numerical solution of the DS equations. This expression for $g(E)$ is also convenient for the evaluation of \eq{Sigmap} as
\be
g'(E)=-\frac{Z}{(E^+-m)^2} + \frac{1}{\pi} \int_{m+m_\pi}^\infty \frac{\mbox{Im} g(\omega)}{(E^+-\w)^2} d\w. 
\ee

\subsection{Separable potential model}
To keep this model as simple as possible, we choose a separable form for the partial wave potential $v$:
\be
v(k',k) =  h(k')\, \lambda\, h(k),
\ee
where $h(k)$ is a phenomenological form factor and $\lambda$ specifies the sign of the potential (for the $P_{11}$ partial wave under consideration here, $\lambda=-1$). In this case \eq{lse} can be solved algebraically, giving also a separable form for the background $\pi N$ t matrix:
\be
t^{(1)}(k',k,E) = h(k')\, \tau(E)\, h(k)
\ee
where
\be
\tau(E) = \left[ 1 - \lambda  \int_0^\infty dq\, q^2\, h(q) g(E-\w_q) h(q)\right]^{-1}\lambda.
\ee
\vspace{-2mm}

\noindent Defining the four dressing terms
\begin{align}
\Sigma_{ij}(E) &=  \int_0^\infty dq\, q^2\, \phi_i(q) g(E-\w_q) \phi_j(q), \hspace{1cm}  (i,j=1,2) \eqn{Sigma}
\end{align}
where $\phi_1(q)\equiv f_0(q)$ and  $\phi_2(q)\equiv h(q)$, 
the DS equations can be conveniently expressed as the set of three equations
\begin{subequations} \eqn{dsg}
\begin{align}
 g(E) &= \frac{1}{E^+ -m-\Sigma(E)+\Sigma(m)} \eqn{gexp} \\[2mm]
 \Sigma(E) &= \Sigma_{11}(E) + \Sigma_{12}(E) \tau(E) \Sigma_{21}(E) \\[2mm]
\tau(E) &= \left[ 1 - \lambda \Sigma_{22}(E) \right]^{-1}\lambda,
\end{align}
\end{subequations}
which determine the dressed nucleon propagator $g(E)$, together with the additional two equations
\begin{subequations} \eqn{dst}
\begin{align}
t(k',k,E)&=f(k,E) g(E) \f(k,E) +h(k')\tau(E)h(k)  \\[2mm]
 f(k,E) &= f_0(k) +  h(k)\tau(E) \Sigma_{21}(E),
\end{align}
\end{subequations}
that determine the consequent dressed $\pi N N$ vertex $f$ and $\pi N$ t matrix $t$.

\subsection{Numerical Procedure}

In our approach, modelling $\pi N$ scattering with \eqs{dspw} begins by choosing parametrized analytic functions for the form factors  $f_0(k)$ and $h(k)$. These form factors need to fulfil the requirement of providing a momentum cutoff that ensures finite values for the integrals defining the $\Sigma$ functions of \eq{Sigma}, and they must behave linearly with $k$ in the limit of low momenta in order to be consistent with the $l=1$ nature of a  $P_{11}$ partial wave amplitude. We shall follow a previous work where separable potentials were used to model $\pi N$ scattering, and choose the following analytic forms\cite{McLeod:1984cu}
\begin{subequations}  \eqn{ff}
\begin{align}
f_0(k)=& \frac{k\, C_0}{\sqrt{\w_\pi(k)}} \frac{1}{(k^2+\Lambda^2)^{n_0}} ,\\[2mm]
h(k) =& \frac{k\, C_1}{\sqrt{\w_\pi(k)}}\left[ \frac{1}{(k^2+\B_1^2)^{n_1}} + \frac{C_2 k^{2n_2}}{(k^2+\B_2^2)^{n_3}}\right],
\end{align}
\end{subequations}
where $C_0, C_1, C_2, \B_1, \B_2, \Lambda$ are free parameters, and the powers $n_0, n_1, n_2, n_3$ are integers that can be chosen to change the functional form of the form factors.

For any given set of parameters, the first task is to solve the DS equations in the form of \eqs{dsg} for the dressed nucleon propagator $g(E)$. We do this by following an iterative procedure where an approximation to $g(E)$ is used in \eq{Sigma} to calculate all the functions $\Sigma_{ij}(E)$, which are then used to calculate a new (and hopefully more accurate) version of $g(E)$ using \eqs{dsg}. The process is repeated until convergence for $g(E)$ is achieved. By construction, the resulting propagator $g(E)$ satisfied the DS equations and can be then used to generate the dressed vertex $f$ and the $\pi N$ t matrix $t$ using \eqs{dst}.

\subsubsection{Technical Aspects}

To carry out the integral in \eq{Sigma} numerically, we use Gaussian quadratures, and to avoid the singularity coming from the pole of $g(E)$, we rotate the integration contour from the positive real axis, into the 4th quadrant of the complex $q$ plane. However, a practical problem remains in carrying out these integrals because in order to generate a propagator $g(E)$ at any iteration, one needs to know the previous iteration's propagators $g(E-\w_{q_i})$  for each of the rotated quadrature points $q_i$. Thus the number of energies at which one needs to know $g$ quickly escalates as the iteration proceeds. To get around this problem, we make use of the fact that the dressed nucleon propagator $g(E)$, at each step of the iteration, satisfies the dispersion relation of \eq{disp}. This allows us to evaluate \eq{Sigma} as
\begin{align}
\Sigma_{ij}&(E) =  \int_0^\infty dq\, q^2\, \frac{Z \phi_i(q) \phi_j(q) }{E^+-\w_q-m} \nn
&- \frac{1}{\pi} \int_{m+m_\pi}^\infty \mbox{Im} g(\omega) \int_0^\infty dq\, q^2\,\frac{\phi_i(q) \phi_j(q) }{E^+-\w_q-\w} d\w \eqn{Sigma1}
\end{align}
which requires knowledge of $g(E)$ only at a number of fixed values of $E$ corresponding to the Gaussian integration points $\w_i$ used to evaluate the $\w$ integral in \eq{Sigma1}. The iterative process thus proceeds according to the following steps:
\begin{enumerate}
\item For any given set of form factor parameters, begin the iteration by generating the ``non-Dyson'' dressed nucleon propagator $g^{(0)}(E)$ defined by \eqs{dsg} but where the nucleon in the $\pi N$ propagator used in the dressing terms $\Sigma_{ij}$, is not explicitly dressed; that is, by using 
\begin{align}
\Sigma_{ij}(E) & \rightarrow  \int_0^\infty dq\, q^2\,\frac{ \phi_i(q) \phi_j(q)}{E^+-\w_q-m}. \eqn{Sigma0}
\end{align}
It is just this $g^{(0)}(E)$ that has been used in previous works\cite{Afnan:1984cv,McLeod:1984cu} to model $\pi N$ scattering.
\item
Having constructed the ``zero'th iteration'' of $g(E)$ as above, we now use this  $g$ in \eq{Sigma1} to generate new dressing terms $ \Sigma_{ij}(E)$.
\item
Using these newly constructed $\Sigma_{ij}(E)$'s in \eqs{dsg} generates the next iteration of the dressed nucleon propagator $g(E)$.
\end{enumerate}

\subsection{Numerical Results}
\begin{figure*}[t] 
   \centering
\raisebox{2.5cm}{(a)}\,\,
  \includegraphics[width=7.5cm]{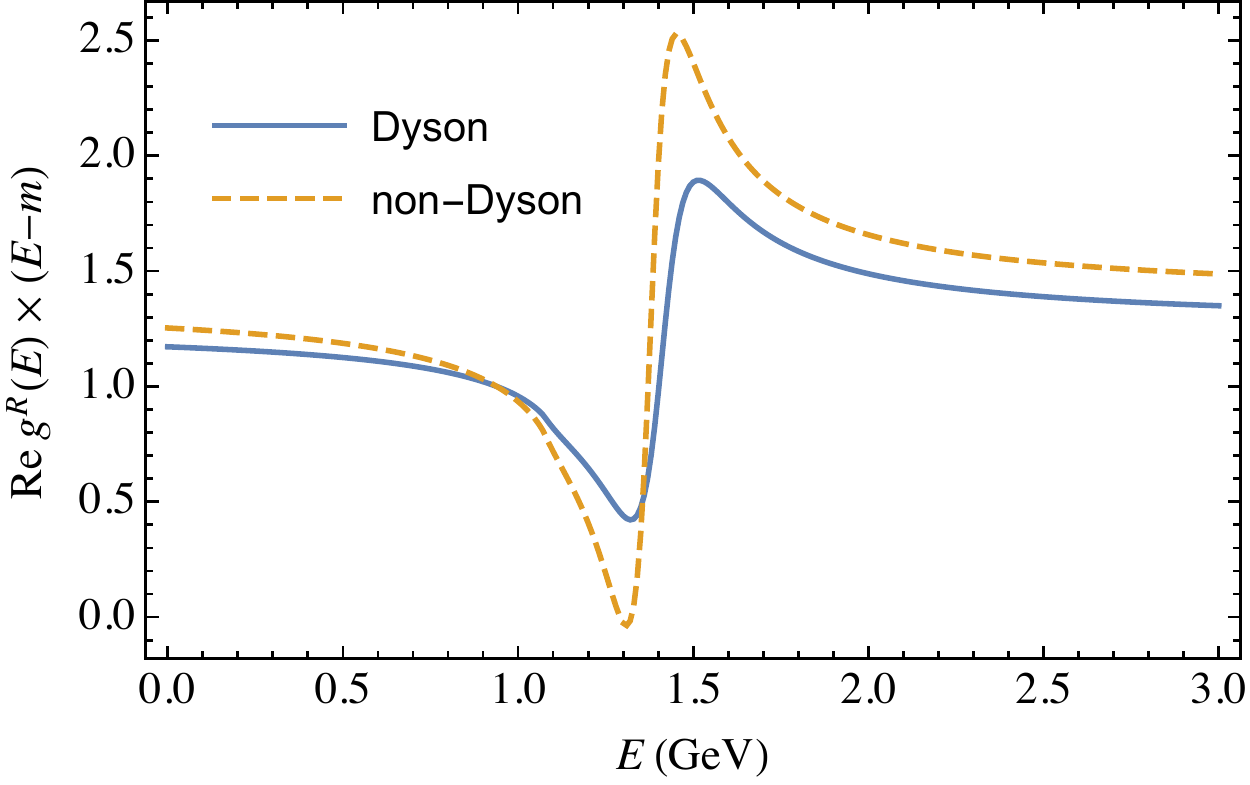} \hspace{2mm}
\raisebox{2.5cm}{(b)}\, \,
\includegraphics[width=7.5cm]{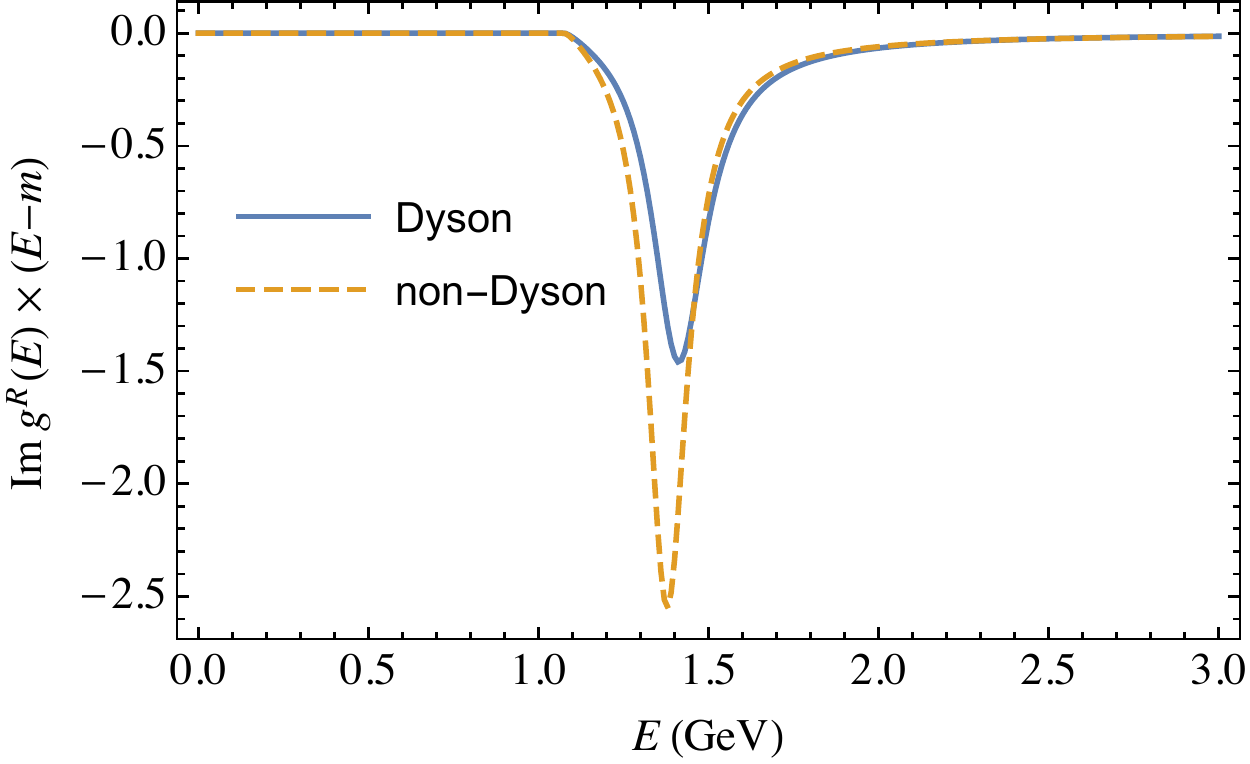} 
   \caption{\fign{G0ratio}
(a) Real and (b) imaginary parts  of $g^R(E) (E-m)$ where $g^R(E)=g(E)/Z$ is the renormalised dressed nucleon propagator. The solid curves are for the Dyson propagator (resulting from the solution of the DS equations), while the dashed curves are for the non-Dyson propagator (resulting from the use of the one-pion-approximation).}
   \label{fig:gg}
\end{figure*}
\begin{table*}[t]
\caption{\label{tab:param} Parameters of four fits (labelled as models $M9$, $M8$, $M7$, and $M6$) to the $P_{11}$ $\pi N$ phase shifts through the solution of the Dyson-Schwinger equations of \eqs{dspw}. The first 9 parameters refer to the form factors of \eq{ff} while $m_0$ is the bare nucleon mass and $Z$ is the nucleon wave function renormalisation constant.}
\begin{ruledtabular}
\begin{tabular}{ccccccccccccc}
  &  &  &  & $\Lambda$ & $\B_1$& $\B_2$& $C_0$& $C_1$& $C_2$ & $m_0$ & $Z$   \\ 
  Potential & $n_0$ & $n_2$ & $n_3$ & $\fm$ & $\fm$& $\fm$& & & & $\fm$ &  \\[1mm] \hline
  $M9$ & 1 & 2 & 3 &  1.71991 &1.14162 & 1.88154 & 0.64225 & 0.23808 & 8.9632 & 5.00289 & 0.90054 \\
$M8$ & 1 & 2 & 3 &  2.72329 &1.26878 & 1.78328 & 1.1778 & 0.32923 & 6.2150 & 5.33579 & 0.79369 \\
$M7$ & 1 & 2 & 3 & 4.04994 & 1.4176 & 1.77246 & 1.7827 & 0.42271 & 4.8618 & 5.69286 & 0.69481  \\
$M6$ & 1 & 2 & 3 & 10.8023 & 1.62718 & 1.8684 & 4.7174 & 0.58264 & 3.5638 & 6.29540  & 0.60027 
\end{tabular}
\end{ruledtabular}
\end{table*}

After constructing the dressed nucleon propagator $g(E)$ as prescribed by the 3 steps outlined above, we now repeat steps 2 and 3 over and over, thus generating successive iterations of $g(E)$, denoted as $g^{(1)}(E), g^{(2)}(E), g^{(3)}(E),\ldots$ until the values of $g^{(r)}(E)$  converge according to the criterion $\left|[g^{(r)}(\w_i)-g^{(r-1)}(\w_i)]/g^{(r)}(\w_i)\right| < \epsilon$ for all $\w$  integration points $\w_i$, where $\epsilon$ is some chosen tolerance value. We have found that the iterated dressed propagators $g^{(r)}(E)$ converge for all considered models using a convergence tolerance of $\epsilon = 10^{-4}$, and that correspondingly, the resulting numerical values of the converged $g(E)$ functions are stable to at least 5 significant figures with respect to variations in the number of quadrature points used for all integrals, and in the contour rotation angle used for all the $q$ momentum integrals.

With the Dyson-Schwinger equations of \eqs{dspw} solved in this way, it is interesting to compare the resulting fully dressed ``Dyson'' nucleon propagator $g(E)$ with the ``non-Dyson'' one where the coupled equations of \fig{fig:eq} are solved in the ``one-pion approximation'' where nucleon dressing in $\pi N$ states is neglected. We present this comparison in \fig{fig:gg} for the case where the parameters of the input bare $\pi N N$ vertex and background $\pi N$ potential are those of model  $M1$ in Ref.\ \cite{McLeod:1984cu}. For ease of comparison, we have plotted the corresponding  real and imaginary values of $(E-m) g(E)/Z$, being the renormalised nucleon propagators with the nucleon pole term factored out. As can be seen, there is a substantial difference between the two propagators, suggesting the importance of retaining nucleon dressing in $\pi N$ states.

To obtain a variety of models of nucleon dressing, we have carried out fits to the WI08 $P_{11}$ $\pi N$ phase shifts \cite{Workman:2012hx} (for pion laboratory energies up to 390 MeV) by using  the functional forms of \eqs{ff} for a number of choices of the integers $n_0$ - $n_3$, and for a range of cutoff values $\Lambda$ for the bare $\pi NN$ vertex function $f_0(k)$. Each such fit was constrained to reproduce the $\pi NN$ coupling constant $f^2_{\pi NN} = 0.079$ in the way described in Ref.\ \cite{McLeod:1984cu} The parameters of four such fits are given  in Table \ref{tab:param} with the corresponding values of $(E-m) g(E)/Z$ plotted in \fig{fig:gmodels}. Unsurprisingly, the large number of parameters in this model allows one to fit $\pi N$ data equally well for a wide range of cutoff parameters $\Lambda$. Although this flexibility of the model can be viewed as one of its weaknesses, it does allow one to accommodate the wide variety of $\pi N N$ vertex cutoffs, in the range $300 < \Lambda < 2200$ MeV,  used in the literature \cite{Cohen:1986ux,Gross:1992tj,Schutz:1994ue,Sato:1996gk,Bockmann:1999nu,Pascalutsa:2000bs,Afnan:2002we,Oettel:2002cw,Kamano:2013iva,Ronchen:2012eg,Skawronski:2018yhu}. 

Finally we show that \eqs{dspw}, which use the fully dressed Dyson propagators, are able to be used to fit all $s$\,- and $p$\,-wave $\pi N$ phase shifts for pion lab kinetic energies in the range $0 < T_{lab} < 390$ MeV. To demonstrate this explicitly, we have chosen  the $M7$ model of Table \ref{tab:param} whose cutoff parameter is $\Lambda=800$ MeV, a value suggested by an investigation of Quantum Chromodynamic sum rules \cite{Meissner:1995ra}. For the non-$P_{11}$ partial wave $\pi N$ potentials, we use the separable forms of Thomas \cite{Thomas:1976px} whose form factors are parametrised as 
\begin{subequations}\eqn{thomas}
\be
h(k) = \frac{S_1}{\A_1^2+k^2} +  \frac{S_2}{\A_2^2+k^2}, \eqn{thomas1}
\ee
for $s$-waves and 
\be
h(k) = \frac{S_1 k}{(\A_1^2+k^2)^2} +  \frac{S_2k^3}{(\A_2^2+k^2)^2}, \eqn{thomas2}
\ee
\end{subequations}
for $p$\,-waves. These form factors  were used by Thomas to describe pion-deuteron scattering in a calculation using semi-relativistic kinematics. \cite{Thomas:1976px}
\begin{figure*}[t] 
   \centering
\raisebox{2.5cm}{\scriptsize(a)}\,
  \includegraphics[width=7.5cm]{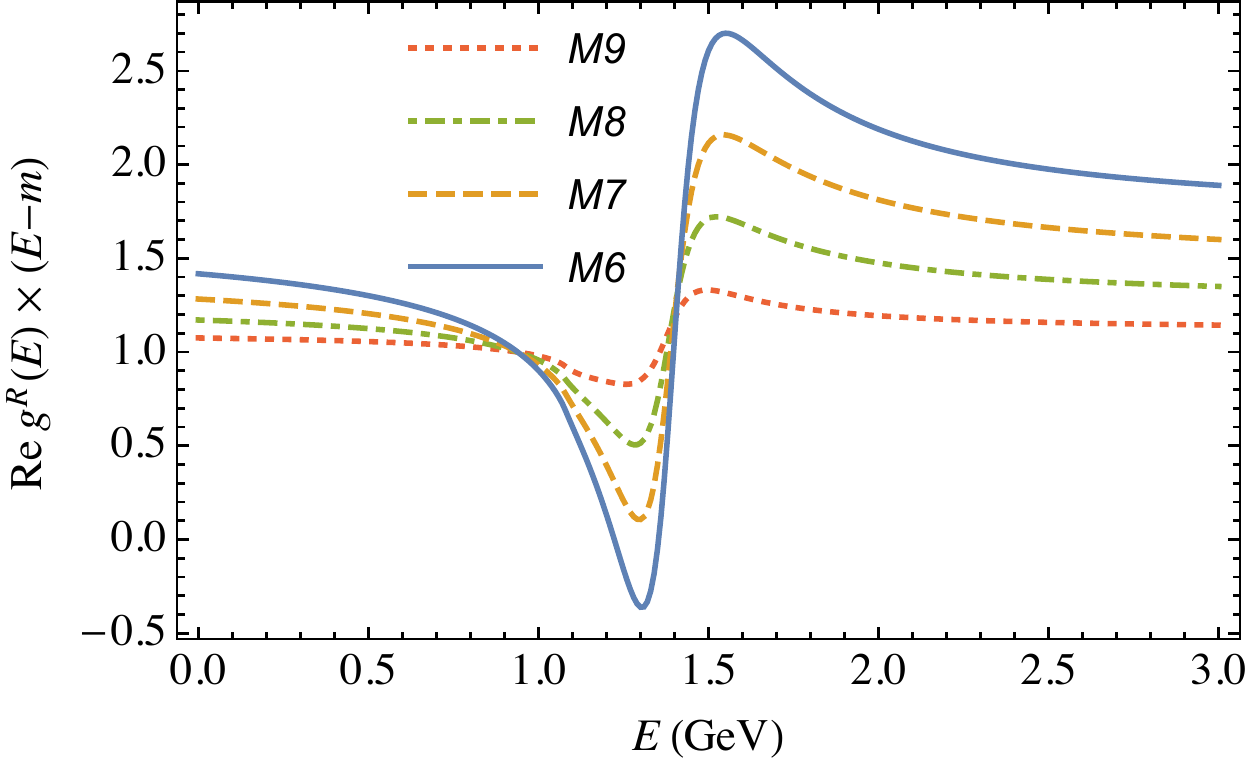} \hspace{0mm}
\raisebox{2.5cm}{\scriptsize(b)}\, 
\includegraphics[width=7.5cm]{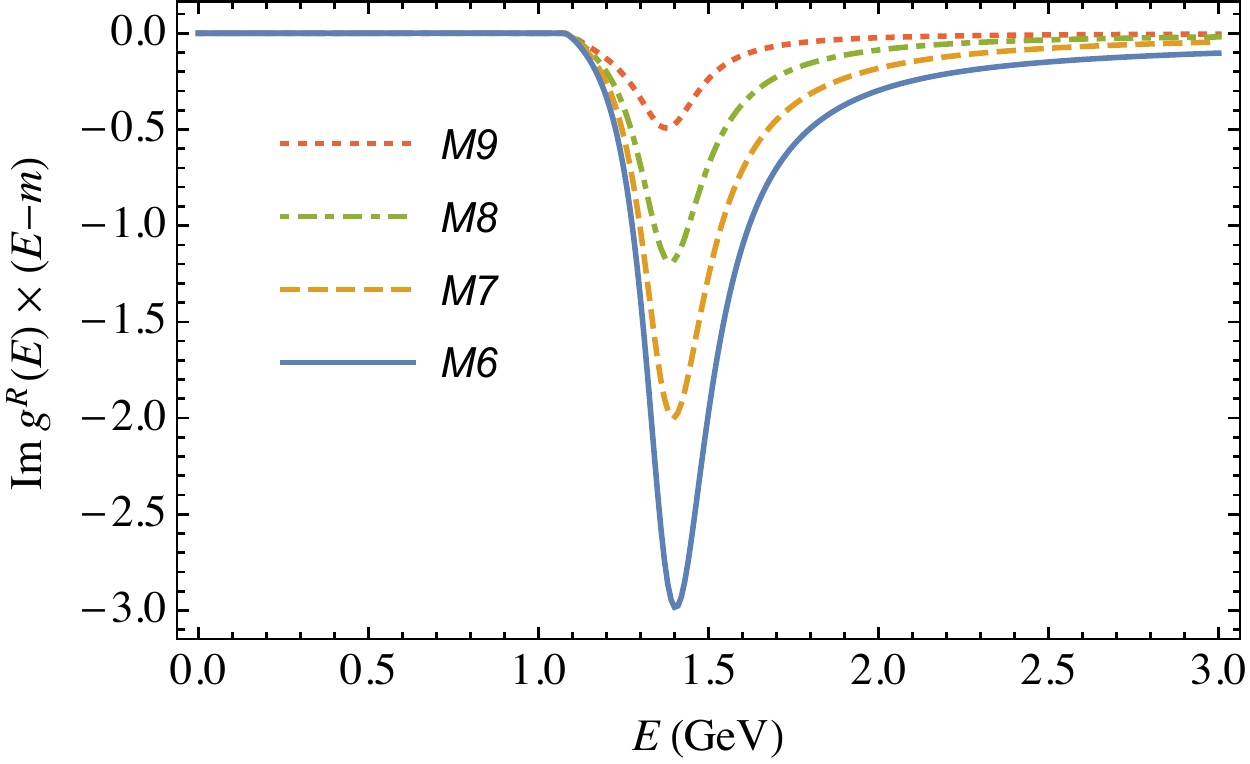} 
   \caption{\fign{fig:gmodels} (a) Real and  (b) imaginary parts of $g^R(E) (E-m)$ where $g^R(E)$ is the renormalized dressed Dyson nucleon propagator,  for the four models specified in Table \ref{tab:param}.}
\end{figure*}
\begin{figure*}[t] 
   \centering
  \includegraphics[width=6.7cm]{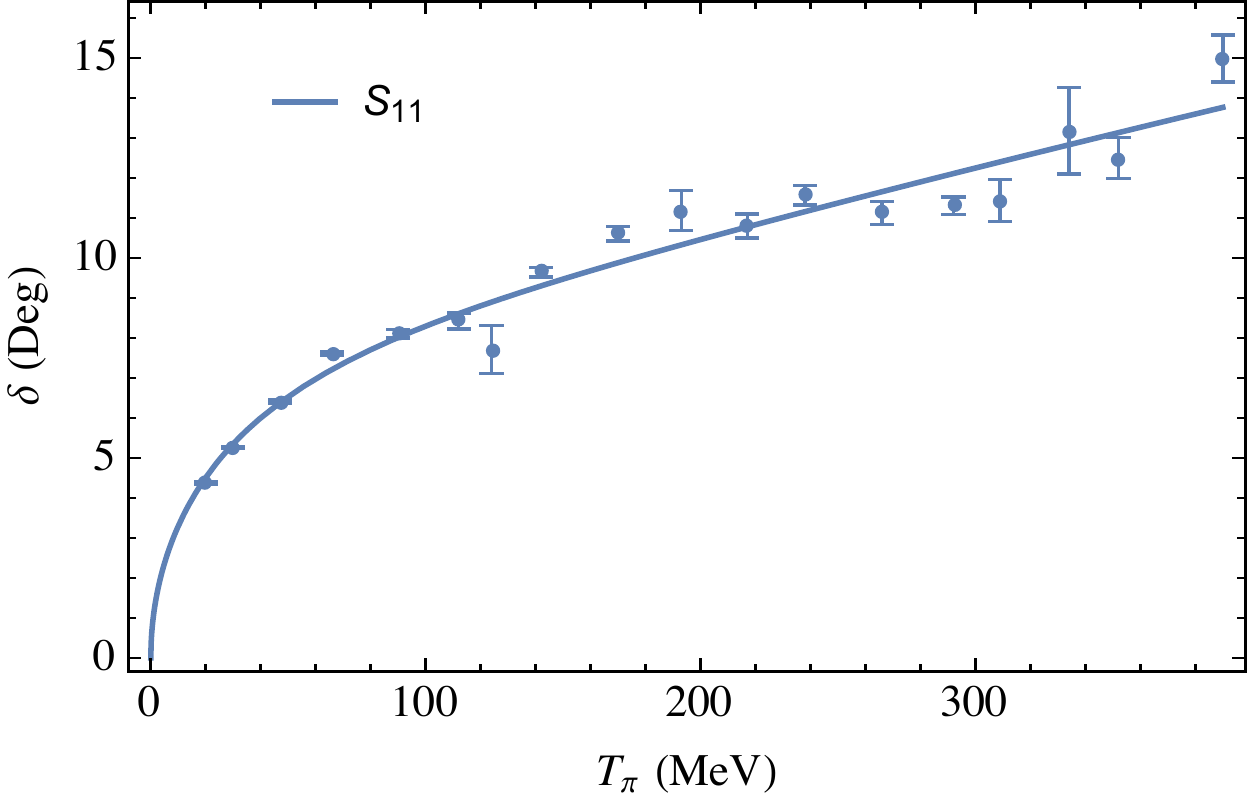} \hspace{5mm}
\includegraphics[width=6.7cm]{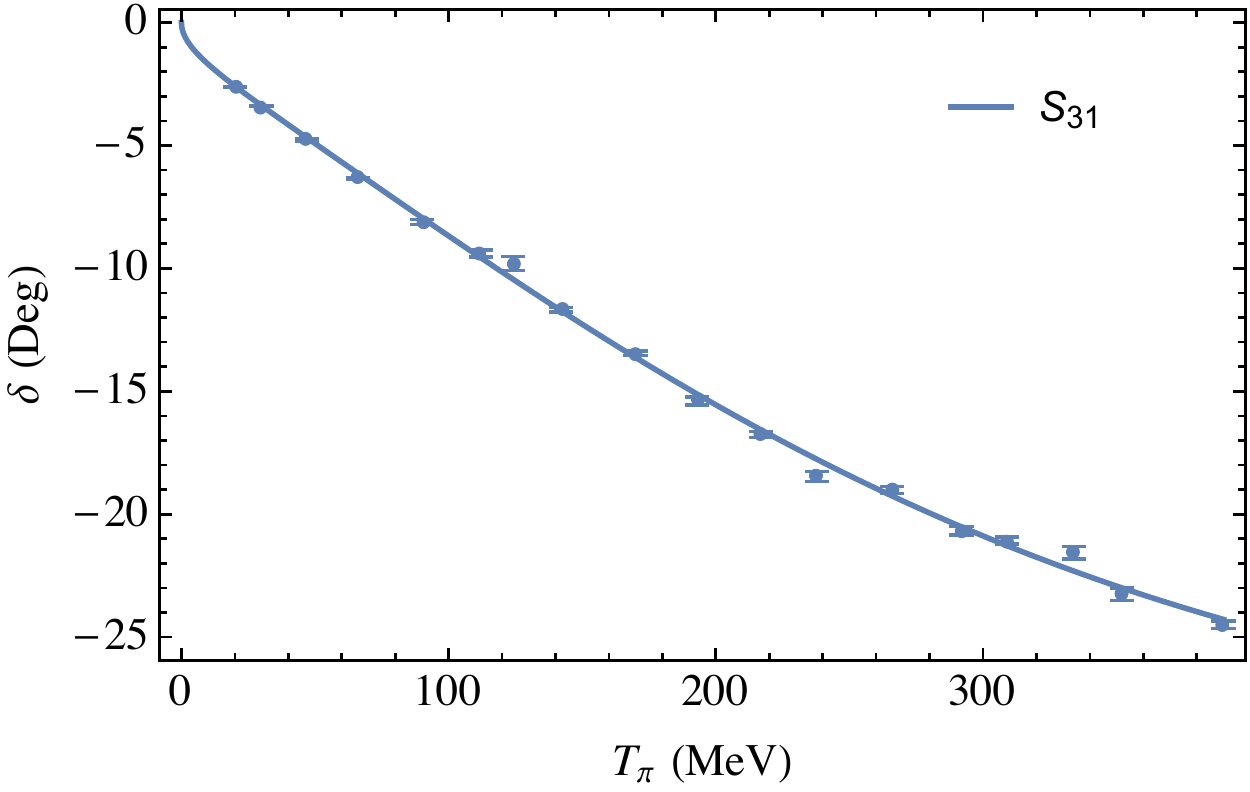} \\
  \includegraphics[width=6.7cm]{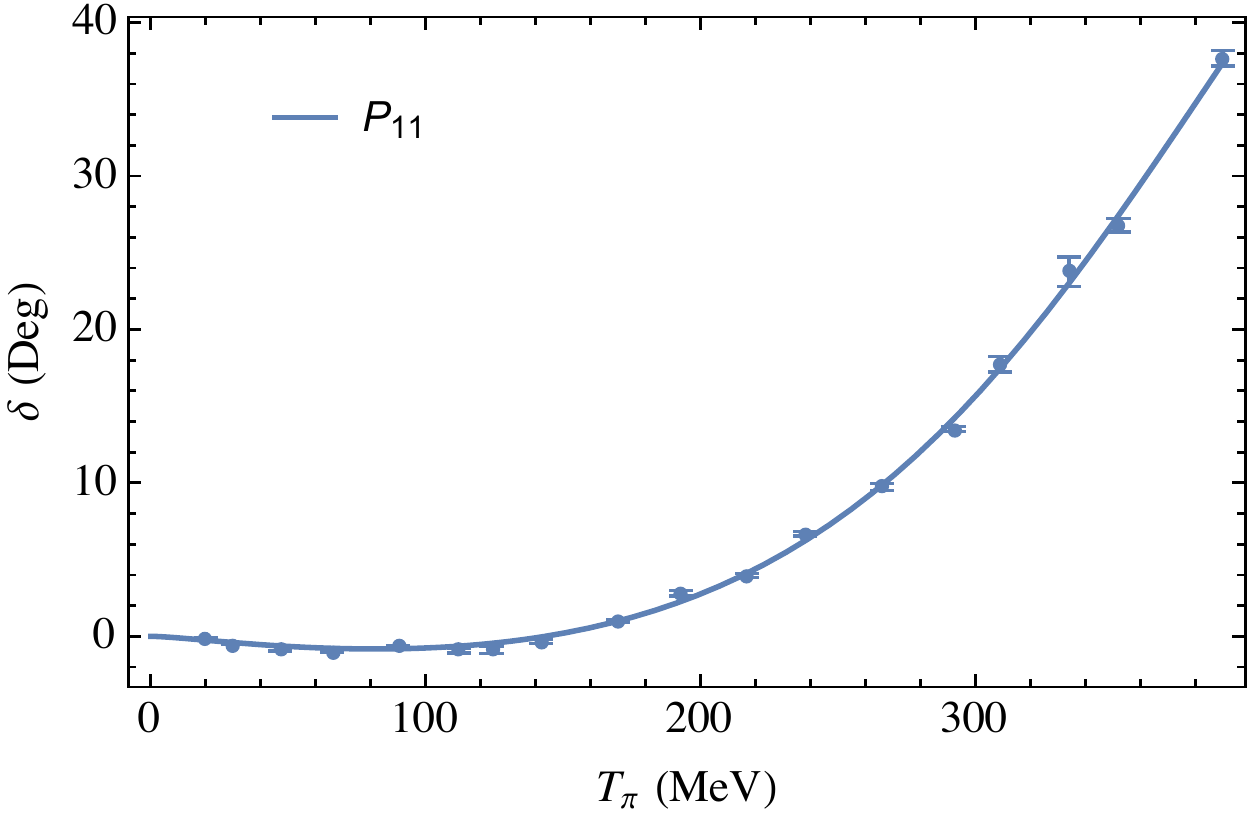} \hspace{5mm}
\includegraphics[width=6.7cm]{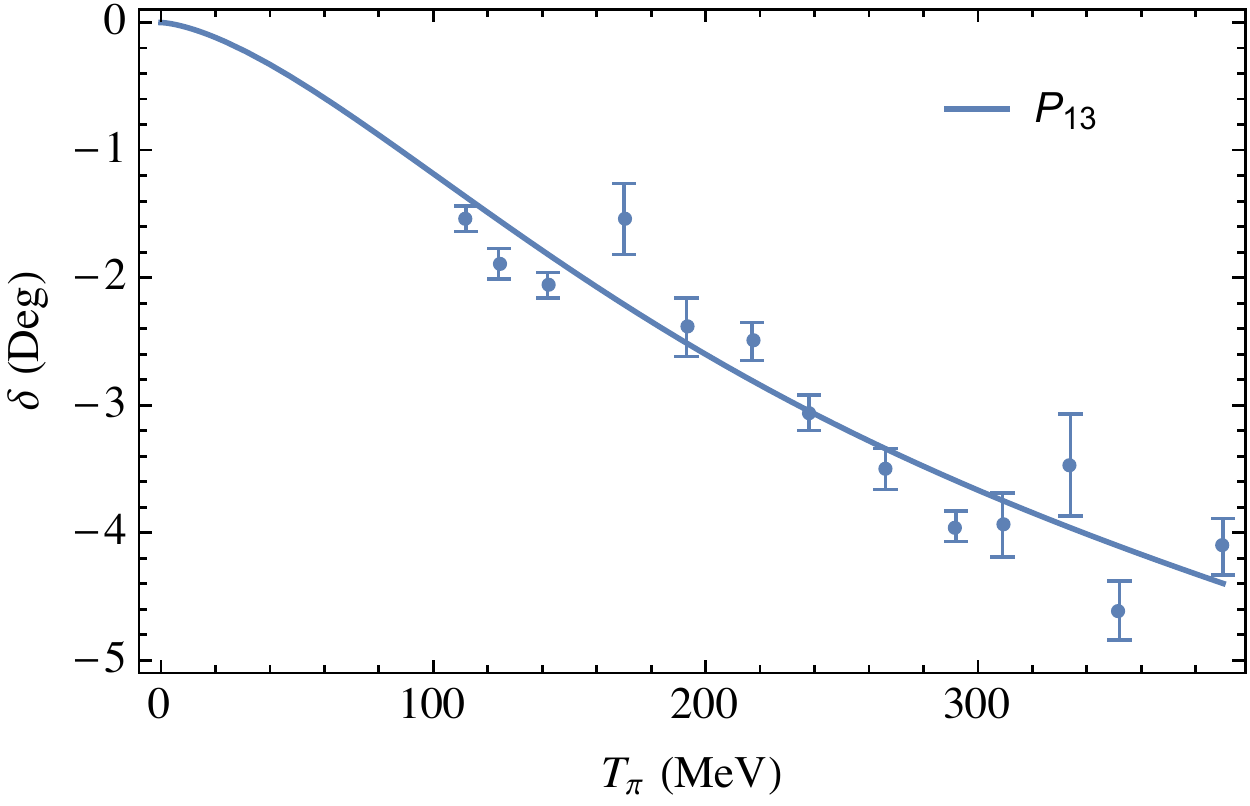} \\
  \includegraphics[width=6.7cm]{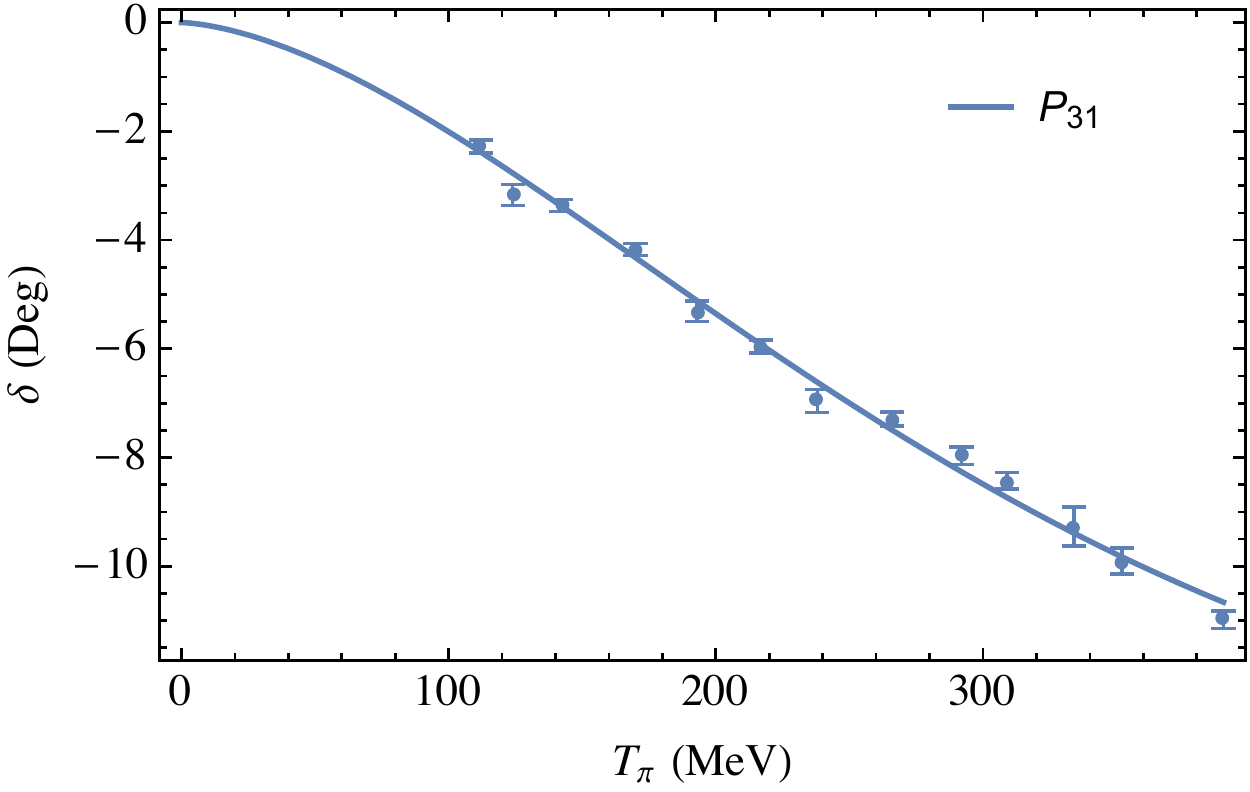} \hspace{5mm}
\includegraphics[width=6.7cm]{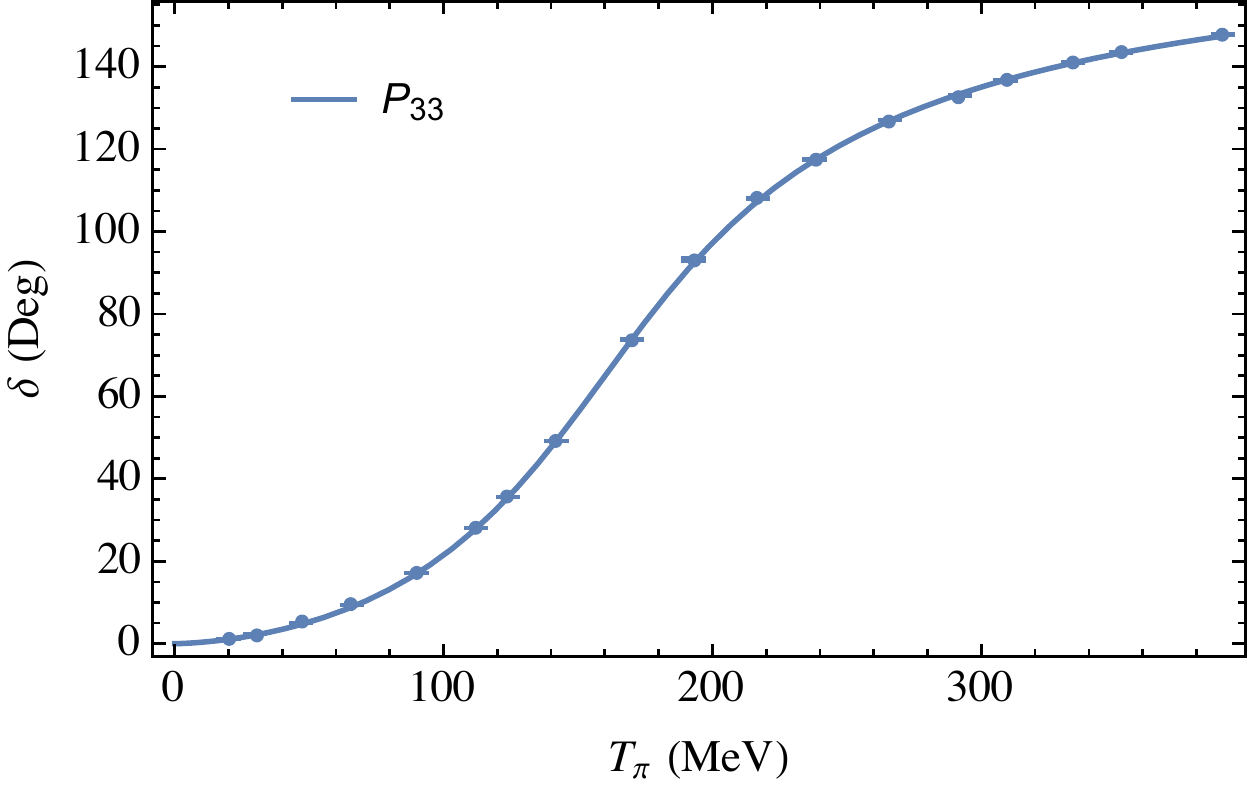}
   \caption{\fign{fig:phases} Fits to the $s$\,- and $p$\,-wave $\pi N$ phase shifts resulting from the solution of the DS equations,  \eqs{dspw}, where the dressed Dyson nucleon propagator $g(E)$ corresponds to the $M7$ model of Table \ref{tab:param}. The $P_{11}$ phased shift fit is the one using the $M7$ model, while all non-$P_{11}$ phase shifts fits are specified by the separable form factor of \eq{thomas} with corresponding parameters given in Table \ref{tab:phases}.}
\end{figure*}

Our partial wave phase shift fits using the $M7$ model for the Dyson propagator $g(E)$ are shown in \fig{fig:phases} with the corresponding parameters listed (under the rows labelled $M7$) in Table  \ref{tab:phases}. Equally good fits to all the phase shifts can be obtained using the other models for $g(E)$ ($M9$, $M8$ and $M6$) with the corresponding parameters being given in Table \ref{tab:param} for the $P_{11}$ and  Table \ref{tab:phases} for the other partial waves.

In summary, we have presented a simple but field-theoretically complete model of $\pi N$ scattering that demonstrates a number of basic features of RQFT such as particle absorption and creation, self-energy, renormalization, and the use of Dyson-Schwinger equations, but without the inherent complexity of a relativistically covariant approach. To do this we used the non-relativistic framework of TOPT, and exploited the simplicity of separable potentials to demonstrate how the DS equations can be solved in order to describe experimental data. It is worth noting that the model presented, together with the resulting fits to $\pi N$ phase shifts as recorded in \fig{fig:phases} and Tables \ref{tab:param} and \ref{tab:phases}, can be of direct practical use in models of pion-multi-nucleon systems, such as the convolution model of the $\pi NN$ system \cite{Kvinikhidze:1992sv, Blankleider:1993dc}, where a complete dressing of all nucleons is required.

 \begin{table}[H]
\caption{\label{tab:phases} Parameters of fits to the $s$\,- and $p$\,-wave $\pi N$ phase shifts (other than $P_{11}$) for each of the 
four models ($M9$, $M8$, $M7$, and $M6$) for the dressed Dyson nucleon propagator $g(E)$  used in the coupled $\pi N$ equations of \eqs{dspw}. The parameters refer to the form factors of \eqs{thomas}. The $s$\,-wave strengths are in $\text{fm}^{-1}$, for $p$\,-waves $S_2$ is dimensionless and $S_1$ is in $\text{fm}^{-1}$.}
\begin{ruledtabular}
\begin{tabular}{ccccccc}
           $\pi N$               & $\lambda$             & $P_{11}$  & $S_1$   & $\alpha_1$       & $S_2$     & $\alpha_2$       \\
      Partial wave                    &                       &       model                &         & $(\text{fm}^{-1})$ &       & $(\text{fm}^{-1})$ \\[1mm] \hline
\multirow{4}{*}{$S_{11}$} & \multirow{4}{*}{$-1$} & $M9$                    & $-11.557$ & 10.057       & $-0.12483$ & 0.85466           \\[-1mm]
                          &                       & $M8$                                           & $-11.715$ & 9.9173       & $-0.13350$  & 0.85973          \\[-1mm]
                          &                       & $M7$                                           & $-11.818$ & 9.7449       & $-0.14334$  & 0.86536          \\[-1mm]
                          &                       & $M6$                                           & $-12.046$ & 9.6221       & $-0.15518$  & 0.87236          \\[1mm] \hline
\multirow{4}{*}{$S_{31}$} & \multirow{4}{*}{$+1$}   & $M9$                  & 3.8017  & 1.9930           & $-1.0897$  & 1.3266            \\[-1mm]
                          &                       & $M8$                                           & 4.2003  & 2.0120           & $-1.1815$  & 1.3292          \\[-1mm]
                          &                       & $M7$                                           & 4.7214  & 2.0350           & $-1.3090$  & 1.3365          \\[-1mm]
                          &                       & $M6$                                           & 5.4998  & 2.0668           & $-1.5227$ & 1.3574          \\[1mm] \hline
\multirow{4}{*}{$P_{31}$} & \multirow{4}{*}{$+1$}   & $M9$                    & 9.5578  & 2.5488           &           &                    \\[-1mm]
                          &                       & $M8$                                           & 10.339  & 2.5549           &           &                  \\[-1mm]
                          &                       & $M7$                                           & 11.250  & 2.5620           &           &                  \\[-1mm]
                          &                       & $M6$                                           & 12.370  & 2.5704           &           &                  \\[1mm] \hline
\multirow{4}{*}{$P_{13}$} & \multirow{4}{*}{$+1$}   & $M9$                 & 3.6208  & 1.8244           & 1.5209    & 3.2380           \\[-1mm]
                          &                       & $M8$                                           & 3.5755  & 1.8022           & 1.4249    & 3.0757           \\[-1mm]
                          &                       & $M7$                                           & 3.5051  & 1.7801           & 1.3111    & 2.9082           \\[-1mm]
                          &                       & $M6$                                           & 3.3855  & 1.7492           & 1.1932    & 2.7126           \\[1mm] \hline
\multirow{4}{*}{$P_{33}$} & \multirow{4}{*}{$-1$}   & $M9$                 & 0.74876 & 1.7551          & 1.2295    & 4.9512            \\[-1mm]
                          &                       & $M8$                                           & 0.80133 & 1.7535          & 1.2212    & 4.8276           \\[-1mm]
                          &                       & $M7$                                           & 0.86060 & 1.7519          & 1.2126    & 4.7019           \\[-1mm]
                          &                       & $M6$                                           & 0.93047 & 1.7501          & 1.2041    & 4.5665          
\end{tabular}
\end{ruledtabular}
\end{table}
\begin{acknowledgments}

 A.N.K. was supported by the Georgian Shota Rustaveli National Science Foundation (Grant No. FR17-354).

\end{acknowledgments}

\bibliographystyle{apsrev4-2}
\bibliography{/Users/phbb/Physics/Papers/refn.bib}

\begin{thebibliography}{36}%
\makeatletter
\providecommand \@ifxundefined [1]{%
 \@ifx{#1\undefined}
}%
\providecommand \@ifnum [1]{%
 \ifnum #1\expandafter \@firstoftwo
 \else \expandafter \@secondoftwo
 \fi
}%
\providecommand \@ifx [1]{%
 \ifx #1\expandafter \@firstoftwo
 \else \expandafter \@secondoftwo
 \fi
}%
\providecommand \natexlab [1]{#1}%
\providecommand \enquote  [1]{``#1''}%
\providecommand \bibnamefont  [1]{#1}%
\providecommand \bibfnamefont [1]{#1}%
\providecommand \citenamefont [1]{#1}%
\providecommand \href@noop [0]{\@secondoftwo}%
\providecommand \href [0]{\begingroup \@sanitize@url \@href}%
\providecommand \@href[1]{\@@startlink{#1}\@@href}%
\providecommand \@@href[1]{\endgroup#1\@@endlink}%
\providecommand \@sanitize@url [0]{\catcode `\\12\catcode `\$12\catcode
  `\&12\catcode `\#12\catcode `\^12\catcode `\_12\catcode `\%12\relax}%
\providecommand \@@startlink[1]{}%
\providecommand \@@endlink[0]{}%
\providecommand \url  [0]{\begingroup\@sanitize@url \@url }%
\providecommand \@url [1]{\endgroup\@href {#1}{\urlprefix }}%
\providecommand \urlprefix  [0]{URL }%
\providecommand \Eprint [0]{\href }%
\providecommand \doibase [0]{https://doi.org/}%
\providecommand \selectlanguage [0]{\@gobble}%
\providecommand \bibinfo  [0]{\@secondoftwo}%
\providecommand \bibfield  [0]{\@secondoftwo}%
\providecommand \translation [1]{[#1]}%
\providecommand \BibitemOpen [0]{}%
\providecommand \bibitemStop [0]{}%
\providecommand \bibitemNoStop [0]{.\EOS\space}%
\providecommand \EOS [0]{\spacefactor3000\relax}%
\providecommand \BibitemShut  [1]{\csname bibitem#1\endcsname}%
\let\auto@bib@innerbib\@empty
\bibitem [{\citenamefont {Mizutani}\ and\ \citenamefont
  {Koltun}(1977)}]{Mizutani:1977xw}%
  \BibitemOpen
  \bibfield  {author} {\bibinfo {author} {\bibfnamefont {T.}~\bibnamefont
  {Mizutani}}\ and\ \bibinfo {author} {\bibfnamefont {D.~S.}\ \bibnamefont
  {Koltun}},\ }\href {https://doi.org/10.1016/0003-4916(77)90164-6} {\bibfield
  {journal} {\bibinfo  {journal} {Annals Phys.}\ }\textbf {\bibinfo {volume}
  {109}},\ \bibinfo {pages} {1} (\bibinfo {year} {1977})}\BibitemShut {NoStop}%
\bibitem [{\citenamefont {Avishai}\ and\ \citenamefont
  {Mizutani}(1979)}]{Avishai:1979it}%
  \BibitemOpen
  \bibfield  {author} {\bibinfo {author} {\bibfnamefont {Y.}~\bibnamefont
  {Avishai}}\ and\ \bibinfo {author} {\bibfnamefont {T.}~\bibnamefont
  {Mizutani}},\ }\href {https://doi.org/10.1016/0375-9474(79)90398-1}
  {\bibfield  {journal} {\bibinfo  {journal} {Nucl. Phys.}\ }\textbf {\bibinfo
  {volume} {A326}},\ \bibinfo {pages} {352} (\bibinfo {year}
  {1979})}\BibitemShut {NoStop}%
\bibitem [{\citenamefont {Avishai}\ and\ \citenamefont
  {Mizutani}(1980)}]{Avishai:1979nm}%
  \BibitemOpen
  \bibfield  {author} {\bibinfo {author} {\bibfnamefont {Y.}~\bibnamefont
  {Avishai}}\ and\ \bibinfo {author} {\bibfnamefont {T.}~\bibnamefont
  {Mizutani}},\ }\href {https://doi.org/10.1016/0375-9474(80)90038-X}
  {\bibfield  {journal} {\bibinfo  {journal} {Nucl. Phys.}\ }\textbf {\bibinfo
  {volume} {A338}},\ \bibinfo {pages} {377} (\bibinfo {year}
  {1980})}\BibitemShut {NoStop}%
\bibitem [{\citenamefont {Avishai}\ and\ \citenamefont
  {Mizutani}(1981)}]{Avishai:1981dbt}%
  \BibitemOpen
  \bibfield  {author} {\bibinfo {author} {\bibfnamefont {Y.}~\bibnamefont
  {Avishai}}\ and\ \bibinfo {author} {\bibfnamefont {T.}~\bibnamefont
  {Mizutani}},\ }\href {https://doi.org/10.1016/0375-9474(81)90419-X}
  {\bibfield  {journal} {\bibinfo  {journal} {Nucl. Phys.}\ }\textbf {\bibinfo
  {volume} {A352}},\ \bibinfo {pages} {399} (\bibinfo {year}
  {1981})}\BibitemShut {NoStop}%
\bibitem [{\citenamefont {Thomas}\ and\ \citenamefont
  {Rinat}(1979)}]{Thomas:1979iw}%
  \BibitemOpen
  \bibfield  {author} {\bibinfo {author} {\bibfnamefont {A.~W.}\ \bibnamefont
  {Thomas}}\ and\ \bibinfo {author} {\bibfnamefont {A.~S.}\ \bibnamefont
  {Rinat}},\ }\href {https://doi.org/10.1103/PhysRevC.20.216} {\bibfield
  {journal} {\bibinfo  {journal} {Phys. Rev. C}\ }\textbf {\bibinfo {volume}
  {20}},\ \bibinfo {pages} {216} (\bibinfo {year} {1979})}\BibitemShut
  {NoStop}%
\bibitem [{\citenamefont {Afnan}\ and\ \citenamefont
  {Blankleider}(1980{\natexlab{a}})}]{Afnan:1980hp}%
  \BibitemOpen
  \bibfield  {author} {\bibinfo {author} {\bibfnamefont {I.~R.}\ \bibnamefont
  {Afnan}}\ and\ \bibinfo {author} {\bibfnamefont {B.}~\bibnamefont
  {Blankleider}},\ }\href {https://doi.org/10.1103/PhysRevC.22.1638} {\bibfield
   {journal} {\bibinfo  {journal} {Phys. Rev. C}\ }\textbf {\bibinfo {volume}
  {22}},\ \bibinfo {pages} {1638} (\bibinfo {year}
  {1980}{\natexlab{a}})}\BibitemShut {NoStop}%
\bibitem [{\citenamefont {Afnan}\ and\ \citenamefont
  {Blankleider}(1980{\natexlab{b}})}]{Afnan:1979tv}%
  \BibitemOpen
  \bibfield  {author} {\bibinfo {author} {\bibfnamefont {I.~R.}\ \bibnamefont
  {Afnan}}\ and\ \bibinfo {author} {\bibfnamefont {B.}~\bibnamefont
  {Blankleider}},\ }\href@noop {} {\bibfield  {journal} {\bibinfo  {journal}
  {Phys. Lett.}\ }\textbf {\bibinfo {volume} {B93}},\ \bibinfo {pages} {367}
  (\bibinfo {year} {1980}{\natexlab{b}})}\BibitemShut {NoStop}%
\bibitem [{\citenamefont {Blankleider}\ and\ \citenamefont
  {Afnan}(1981)}]{Blankleider:1981yp}%
  \BibitemOpen
  \bibfield  {author} {\bibinfo {author} {\bibfnamefont {B.}~\bibnamefont
  {Blankleider}}\ and\ \bibinfo {author} {\bibfnamefont {I.~R.}\ \bibnamefont
  {Afnan}},\ }\href {https://doi.org/10.1103/PhysRevC.24.1572} {\bibfield
  {journal} {\bibinfo  {journal} {Phys. Rev. C}\ }\textbf {\bibinfo {volume}
  {24}},\ \bibinfo {pages} {1572} (\bibinfo {year} {1981})}\BibitemShut
  {NoStop}%
\bibitem [{\citenamefont {Afnan}\ and\ \citenamefont
  {Blankleider}(1985)}]{Afnan:1985qm}%
  \BibitemOpen
  \bibfield  {author} {\bibinfo {author} {\bibfnamefont {I.~R.}\ \bibnamefont
  {Afnan}}\ and\ \bibinfo {author} {\bibfnamefont {B.}~\bibnamefont
  {Blankleider}},\ }\href@noop {} {\bibfield  {journal} {\bibinfo  {journal}
  {Phys. Rev. C}\ }\textbf {\bibinfo {volume} {32}},\ \bibinfo {pages} {2006}
  (\bibinfo {year} {1985})}\BibitemShut {NoStop}%
\bibitem [{\citenamefont {Rinat}\ and\ \citenamefont
  {Starkand}(1983)}]{Rinat:1982ry}%
  \BibitemOpen
  \bibfield  {author} {\bibinfo {author} {\bibfnamefont {A.~S.}\ \bibnamefont
  {Rinat}}\ and\ \bibinfo {author} {\bibfnamefont {Y.}~\bibnamefont
  {Starkand}},\ }\href {https://doi.org/10.1016/0375-9474(83)90609-7}
  {\bibfield  {journal} {\bibinfo  {journal} {Nucl. Phys.}\ }\textbf {\bibinfo
  {volume} {A397}},\ \bibinfo {pages} {381} (\bibinfo {year}
  {1983})}\BibitemShut {NoStop}%
\bibitem [{\citenamefont {Mizutani}\ \emph {et~al.}(1987)\citenamefont
  {Mizutani}, \citenamefont {Saghai}, \citenamefont {Fayard},\ and\
  \citenamefont {Lamot}}]{Mizutani:1986qt}%
  \BibitemOpen
  \bibfield  {author} {\bibinfo {author} {\bibfnamefont {T.}~\bibnamefont
  {Mizutani}}, \bibinfo {author} {\bibfnamefont {B.}~\bibnamefont {Saghai}},
  \bibinfo {author} {\bibfnamefont {C.}~\bibnamefont {Fayard}},\ and\ \bibinfo
  {author} {\bibfnamefont {G.~H.}\ \bibnamefont {Lamot}},\ }\href
  {https://doi.org/10.1103/PhysRevC.35.667} {\bibfield  {journal} {\bibinfo
  {journal} {Phys. Rev. C}\ }\textbf {\bibinfo {volume} {35}},\ \bibinfo
  {pages} {667} (\bibinfo {year} {1987})}\BibitemShut {NoStop}%
\bibitem [{\citenamefont {Lamot}\ \emph {et~al.}(1987)\citenamefont {Lamot},
  \citenamefont {Perrot}, \citenamefont {Fayard},\ and\ \citenamefont
  {Mizutani}}]{Lamot:1987tf}%
  \BibitemOpen
  \bibfield  {author} {\bibinfo {author} {\bibfnamefont {G.~H.}\ \bibnamefont
  {Lamot}}, \bibinfo {author} {\bibfnamefont {J.~L.}\ \bibnamefont {Perrot}},
  \bibinfo {author} {\bibfnamefont {C.}~\bibnamefont {Fayard}},\ and\ \bibinfo
  {author} {\bibfnamefont {T.}~\bibnamefont {Mizutani}},\ }\href
  {https://doi.org/10.1103/PhysRevC.35.239} {\bibfield  {journal} {\bibinfo
  {journal} {Phys. Rev. C}\ }\textbf {\bibinfo {volume} {35}},\ \bibinfo
  {pages} {239} (\bibinfo {year} {1987})}\BibitemShut {NoStop}%
\bibitem [{\citenamefont {Mizutani}\ \emph {et~al.}(1989)\citenamefont
  {Mizutani}, \citenamefont {Fayard}, \citenamefont {Lamot},\ and\
  \citenamefont {Saghai}}]{Mizutani:1989uw}%
  \BibitemOpen
  \bibfield  {author} {\bibinfo {author} {\bibfnamefont {T.}~\bibnamefont
  {Mizutani}}, \bibinfo {author} {\bibfnamefont {C.}~\bibnamefont {Fayard}},
  \bibinfo {author} {\bibfnamefont {G.~H.}\ \bibnamefont {Lamot}},\ and\
  \bibinfo {author} {\bibfnamefont {B.}~\bibnamefont {Saghai}},\ }\href
  {https://doi.org/10.1103/PhysRevC.40.2763} {\bibfield  {journal} {\bibinfo
  {journal} {Phys. Rev. C}\ }\textbf {\bibinfo {volume} {40}},\ \bibinfo
  {pages} {2763} (\bibinfo {year} {1989})}\BibitemShut {NoStop}%
\bibitem [{\citenamefont {Fayard}\ \emph {et~al.}(1992)\citenamefont {Fayard},
  \citenamefont {Lamot}, \citenamefont {Mizutani},\ and\ \citenamefont
  {Saghai}}]{Fayard:1992ww}%
  \BibitemOpen
  \bibfield  {author} {\bibinfo {author} {\bibfnamefont {C.}~\bibnamefont
  {Fayard}}, \bibinfo {author} {\bibfnamefont {G.~H.}\ \bibnamefont {Lamot}},
  \bibinfo {author} {\bibfnamefont {T.}~\bibnamefont {Mizutani}},\ and\
  \bibinfo {author} {\bibfnamefont {B.}~\bibnamefont {Saghai}},\ }\href
  {https://doi.org/10.1103/PhysRevC.46.118} {\bibfield  {journal} {\bibinfo
  {journal} {Phys. Rev. C}\ }\textbf {\bibinfo {volume} {46}},\ \bibinfo
  {pages} {118} (\bibinfo {year} {1992})}\BibitemShut {NoStop}%
\bibitem [{\citenamefont {Sauer}\ \emph {et~al.}(1985)\citenamefont {Sauer},
  \citenamefont {Sawicki},\ and\ \citenamefont {Furui}}]{Sauer:1984um}%
  \BibitemOpen
  \bibfield  {author} {\bibinfo {author} {\bibfnamefont {P.~U.}\ \bibnamefont
  {Sauer}}, \bibinfo {author} {\bibfnamefont {M.}~\bibnamefont {Sawicki}},\
  and\ \bibinfo {author} {\bibfnamefont {S.}~\bibnamefont {Furui}},\ }\href
  {https://doi.org/10.1143/PTP.74.1290} {\bibfield  {journal} {\bibinfo
  {journal} {Prog. Theor. Phys.}\ }\textbf {\bibinfo {volume} {74}},\ \bibinfo
  {pages} {1290} (\bibinfo {year} {1985})}\BibitemShut {NoStop}%
\bibitem [{\citenamefont {Kvinikhidze}\ and\ \citenamefont
  {Blankleider}(1993{\natexlab{a}})}]{Kvinikhidze:1992sv}%
  \BibitemOpen
  \bibfield  {author} {\bibinfo {author} {\bibfnamefont {A.~N.}\ \bibnamefont
  {Kvinikhidze}}\ and\ \bibinfo {author} {\bibfnamefont {B.}~\bibnamefont
  {Blankleider}},\ }\href {https://doi.org/10.1016/0370-2693(93)90184-J}
  {\bibfield  {journal} {\bibinfo  {journal} {Phys. Lett.}\ }\textbf {\bibinfo
  {volume} {B307}},\ \bibinfo {pages} {7} (\bibinfo {year}
  {1993}{\natexlab{a}})}\BibitemShut {NoStop}%
\bibitem [{\citenamefont {Kvinikhidze}\ and\ \citenamefont
  {Blankleider}(1993{\natexlab{b}})}]{Kvinikhidze:1992em}%
  \BibitemOpen
  \bibfield  {author} {\bibinfo {author} {\bibfnamefont {A.~N.}\ \bibnamefont
  {Kvinikhidze}}\ and\ \bibinfo {author} {\bibfnamefont {B.}~\bibnamefont
  {Blankleider}},\ }\href {https://doi.org/10.1103/PhysRevC.48.25} {\bibfield
  {journal} {\bibinfo  {journal} {Phys. Rev. C}\ }\textbf {\bibinfo {volume}
  {48}},\ \bibinfo {pages} {25} (\bibinfo {year}
  {1993}{\natexlab{b}})}\BibitemShut {NoStop}%
\bibitem [{\citenamefont {Blankleider}\ and\ \citenamefont
  {Kvinikhidze}(1994)}]{Blankleider:1993dc}%
  \BibitemOpen
  \bibfield  {author} {\bibinfo {author} {\bibfnamefont {B.}~\bibnamefont
  {Blankleider}}\ and\ \bibinfo {author} {\bibfnamefont {A.~N.}\ \bibnamefont
  {Kvinikhidze}},\ }\href@noop {} {\bibfield  {journal} {\bibinfo  {journal}
  {Few Body Syst. Suppl.}\ }\textbf {\bibinfo {volume} {7}},\ \bibinfo {pages}
  {294} (\bibinfo {year} {1994})},\ \Eprint
  {https://arxiv.org/abs/nucl-th/9402011} {nucl-th/9402011} \BibitemShut
  {NoStop}%
\bibitem [{\citenamefont {Fetter}\ and\ \citenamefont
  {Walecka}(2003)}]{Fetter}%
  \BibitemOpen
  \bibfield  {author} {\bibinfo {author} {\bibfnamefont {A.~L.}\ \bibnamefont
  {Fetter}}\ and\ \bibinfo {author} {\bibfnamefont {J.~D.}\ \bibnamefont
  {Walecka}},\ }\href@noop {} {\emph {\bibinfo {title} {Quantum Theory of
  Many-Particle Systems}}}\ (\bibinfo  {publisher} {Dover Publications},\
  \bibinfo {year} {2003})\BibitemShut {NoStop}%
\bibitem [{\citenamefont {Thomas}(1976)}]{Thomas:1976px}%
  \BibitemOpen
  \bibfield  {author} {\bibinfo {author} {\bibfnamefont {A.~W.}\ \bibnamefont
  {Thomas}},\ }\href {https://doi.org/10.1016/0375-9474(76)90483-8} {\bibfield
  {journal} {\bibinfo  {journal} {Nucl. Phys.}\ }\textbf {\bibinfo {volume}
  {A258}},\ \bibinfo {pages} {417} (\bibinfo {year} {1976})}\BibitemShut
  {NoStop}%
\bibitem [{\citenamefont {McLeod}\ and\ \citenamefont
  {Afnan}(1985)}]{McLeod:1984cu}%
  \BibitemOpen
  \bibfield  {author} {\bibinfo {author} {\bibfnamefont {R.~J.}\ \bibnamefont
  {McLeod}}\ and\ \bibinfo {author} {\bibfnamefont {I.~R.}\ \bibnamefont
  {Afnan}},\ }\href@noop {} {\bibfield  {journal} {\bibinfo  {journal} {Phys.
  Rev. C}\ }\textbf {\bibinfo {volume} {32}},\ \bibinfo {pages} {222} (\bibinfo
  {year} {1985})}\BibitemShut {NoStop}%
\bibitem [{\citenamefont {Mizutani}\ \emph {et~al.}(1981)\citenamefont
  {Mizutani}, \citenamefont {Fayard}, \citenamefont {Lamot},\ and\
  \citenamefont {Nahabetian}}]{Mizutani:1981cb}%
  \BibitemOpen
  \bibfield  {author} {\bibinfo {author} {\bibfnamefont {T.}~\bibnamefont
  {Mizutani}}, \bibinfo {author} {\bibfnamefont {C.}~\bibnamefont {Fayard}},
  \bibinfo {author} {\bibfnamefont {G.}~\bibnamefont {Lamot}},\ and\ \bibinfo
  {author} {\bibfnamefont {R.}~\bibnamefont {Nahabetian}},\ }\href
  {https://doi.org/10.1103/PhysRevC.24.2633} {\bibfield  {journal} {\bibinfo
  {journal} {Phys. Rev. C}\ }\textbf {\bibinfo {volume} {24}},\ \bibinfo
  {pages} {2633} (\bibinfo {year} {1981})}\BibitemShut {NoStop}%
\bibitem [{\citenamefont {Afnan}\ and\ \citenamefont
  {McLeod}(1985)}]{Afnan:1984cv}%
  \BibitemOpen
  \bibfield  {author} {\bibinfo {author} {\bibfnamefont {I.~R.}\ \bibnamefont
  {Afnan}}\ and\ \bibinfo {author} {\bibfnamefont {R.~J.}\ \bibnamefont
  {McLeod}},\ }\href@noop {} {\bibfield  {journal} {\bibinfo  {journal} {Phys.
  Rev. C}\ }\textbf {\bibinfo {volume} {31}},\ \bibinfo {pages} {1821}
  (\bibinfo {year} {1985})}\BibitemShut {NoStop}%
\bibitem [{\citenamefont {Workman}\ \emph {et~al.}(2012)\citenamefont
  {Workman}, \citenamefont {Arndt}, \citenamefont {Briscoe}, \citenamefont
  {Paris},\ and\ \citenamefont {Strakovsky}}]{Workman:2012hx}%
  \BibitemOpen
  \bibfield  {author} {\bibinfo {author} {\bibfnamefont {R.}~\bibnamefont
  {Workman}}, \bibinfo {author} {\bibfnamefont {R.}~\bibnamefont {Arndt}},
  \bibinfo {author} {\bibfnamefont {W.}~\bibnamefont {Briscoe}}, \bibinfo
  {author} {\bibfnamefont {M.}~\bibnamefont {Paris}},\ and\ \bibinfo {author}
  {\bibfnamefont {I.}~\bibnamefont {Strakovsky}},\ }\href
  {https://doi.org/10.1103/PhysRevC.86.035202} {\bibfield  {journal} {\bibinfo
  {journal} {Phys. Rev. C}\ }\textbf {\bibinfo {volume} {86}},\ \bibinfo
  {pages} {035202} (\bibinfo {year} {2012})},\ \Eprint
  {https://arxiv.org/abs/1204.2277} {arXiv:1204.2277 [hep-ph]} \BibitemShut
  {NoStop}%
\bibitem [{\citenamefont {Cohen}(1986)}]{Cohen:1986ux}%
  \BibitemOpen
  \bibfield  {author} {\bibinfo {author} {\bibfnamefont {T.~D.}\ \bibnamefont
  {Cohen}},\ }\href {https://doi.org/10.1103/PhysRevD.34.2187} {\bibfield
  {journal} {\bibinfo  {journal} {Phys. Rev. D}\ }\textbf {\bibinfo {volume}
  {34}},\ \bibinfo {pages} {2187} (\bibinfo {year} {1986})}\BibitemShut
  {NoStop}%
\bibitem [{\citenamefont {Gross}\ and\ \citenamefont
  {Surya}(1993)}]{Gross:1992tj}%
  \BibitemOpen
  \bibfield  {author} {\bibinfo {author} {\bibfnamefont {F.}~\bibnamefont
  {Gross}}\ and\ \bibinfo {author} {\bibfnamefont {Y.}~\bibnamefont {Surya}},\
  }\href {https://doi.org/10.1103/PhysRevC.47.703} {\bibfield  {journal}
  {\bibinfo  {journal} {Phys. Rev. C}\ }\textbf {\bibinfo {volume} {47}},\
  \bibinfo {pages} {703} (\bibinfo {year} {1993})}\BibitemShut {NoStop}%
\bibitem [{\citenamefont {Schutz}\ \emph {et~al.}(1994)\citenamefont {Schutz},
  \citenamefont {Durso}, \citenamefont {Holinde},\ and\ \citenamefont
  {Speth}}]{Schutz:1994ue}%
  \BibitemOpen
  \bibfield  {author} {\bibinfo {author} {\bibfnamefont {C.}~\bibnamefont
  {Schutz}}, \bibinfo {author} {\bibfnamefont {J.~W.}\ \bibnamefont {Durso}},
  \bibinfo {author} {\bibfnamefont {K.}~\bibnamefont {Holinde}},\ and\ \bibinfo
  {author} {\bibfnamefont {J.}~\bibnamefont {Speth}},\ }\href
  {https://doi.org/10.1103/PhysRevC.49.2671} {\bibfield  {journal} {\bibinfo
  {journal} {Phys. Rev.}\ }\textbf {\bibinfo {volume} {C49}},\ \bibinfo {pages}
  {2671} (\bibinfo {year} {1994})}\BibitemShut {NoStop}%
\bibitem [{\citenamefont {Sato}\ and\ \citenamefont {Lee}(1996)}]{Sato:1996gk}%
  \BibitemOpen
  \bibfield  {author} {\bibinfo {author} {\bibfnamefont {T.}~\bibnamefont
  {Sato}}\ and\ \bibinfo {author} {\bibfnamefont {T.~S.~H.}\ \bibnamefont
  {Lee}},\ }\href@noop {} {\bibfield  {journal} {\bibinfo  {journal} {Phys.
  Rev.}\ }\textbf {\bibinfo {volume} {C54}},\ \bibinfo {pages} {2660} (\bibinfo
  {year} {1996})},\ \Eprint {https://arxiv.org/abs/nucl-th/9606009}
  {nucl-th/9606009} \BibitemShut {NoStop}%
\bibitem [{\citenamefont {Bockmann}\ \emph {et~al.}(1999)\citenamefont
  {Bockmann}, \citenamefont {Hanhart}, \citenamefont {Krehl}, \citenamefont
  {Krewald},\ and\ \citenamefont {Speth}}]{Bockmann:1999nu}%
  \BibitemOpen
  \bibfield  {author} {\bibinfo {author} {\bibfnamefont {R.}~\bibnamefont
  {Bockmann}}, \bibinfo {author} {\bibfnamefont {C.}~\bibnamefont {Hanhart}},
  \bibinfo {author} {\bibfnamefont {O.}~\bibnamefont {Krehl}}, \bibinfo
  {author} {\bibfnamefont {S.}~\bibnamefont {Krewald}},\ and\ \bibinfo {author}
  {\bibfnamefont {J.}~\bibnamefont {Speth}},\ }\href
  {https://doi.org/10.1103/PhysRevC.60.055212} {\bibfield  {journal} {\bibinfo
  {journal} {Phys. Rev.}\ }\textbf {\bibinfo {volume} {C60}},\ \bibinfo {pages}
  {055212} (\bibinfo {year} {1999})},\ \Eprint
  {https://arxiv.org/abs/nucl-th/9905043} {arXiv:nucl-th/9905043} \BibitemShut
  {NoStop}%
\bibitem [{\citenamefont {Pascalutsa}\ and\ \citenamefont
  {Tjon}(2000)}]{Pascalutsa:2000bs}%
  \BibitemOpen
  \bibfield  {author} {\bibinfo {author} {\bibfnamefont {V.}~\bibnamefont
  {Pascalutsa}}\ and\ \bibinfo {author} {\bibfnamefont {J.~A.}\ \bibnamefont
  {Tjon}},\ }\href {https://doi.org/10.1103/PhysRevC.61.054003} {\bibfield
  {journal} {\bibinfo  {journal} {Phys. Rev. C}\ }\textbf {\bibinfo {volume}
  {61}},\ \bibinfo {pages} {054003} (\bibinfo {year} {2000})},\ \Eprint
  {https://arxiv.org/abs/nucl-th/0003050} {arXiv:nucl-th/0003050} \BibitemShut
  {NoStop}%
\bibitem [{\citenamefont {Afnan}\ and\ \citenamefont
  {Lahiff}(2003)}]{Afnan:2002we}%
  \BibitemOpen
  \bibfield  {author} {\bibinfo {author} {\bibfnamefont {I.~R.}\ \bibnamefont
  {Afnan}}\ and\ \bibinfo {author} {\bibfnamefont {A.~D.}\ \bibnamefont
  {Lahiff}},\ }\href {https://doi.org/10.1140/epja/i2002-10221-7} {\bibfield
  {journal} {\bibinfo  {journal} {Eur. Phys. J.}\ }\textbf {\bibinfo {volume}
  {A18}},\ \bibinfo {pages} {301} (\bibinfo {year} {2003})},\ \Eprint
  {https://arxiv.org/abs/nucl-th/0210027} {arXiv:nucl-th/0210027 [nucl-th]}
  \BibitemShut {NoStop}%
\bibitem [{\citenamefont {Oettel}\ and\ \citenamefont
  {Thomas}(2002)}]{Oettel:2002cw}%
  \BibitemOpen
  \bibfield  {author} {\bibinfo {author} {\bibfnamefont {M.}~\bibnamefont
  {Oettel}}\ and\ \bibinfo {author} {\bibfnamefont {A.~W.}\ \bibnamefont
  {Thomas}},\ }\href {https://doi.org/10.1103/PhysRevC.66.065207} {\bibfield
  {journal} {\bibinfo  {journal} {Phys. Rev. C}\ }\textbf {\bibinfo {volume}
  {66}},\ \bibinfo {pages} {065207} (\bibinfo {year} {2002})},\ \Eprint
  {https://arxiv.org/abs/nucl-th/0203073} {arXiv:nucl-th/0203073 [nucl-th]}
  \BibitemShut {NoStop}%
\bibitem [{\citenamefont {Kamano}\ \emph {et~al.}(2013)\citenamefont {Kamano},
  \citenamefont {Nakamura}, \citenamefont {Lee},\ and\ \citenamefont
  {Sato}}]{Kamano:2013iva}%
  \BibitemOpen
  \bibfield  {author} {\bibinfo {author} {\bibfnamefont {H.}~\bibnamefont
  {Kamano}}, \bibinfo {author} {\bibfnamefont {S.~X.}\ \bibnamefont
  {Nakamura}}, \bibinfo {author} {\bibfnamefont {T.~S.~H.}\ \bibnamefont
  {Lee}},\ and\ \bibinfo {author} {\bibfnamefont {T.}~\bibnamefont {Sato}},\
  }\href {https://doi.org/10.1103/PhysRevC.88.035209} {\bibfield  {journal}
  {\bibinfo  {journal} {Phys. Rev. C}\ }\textbf {\bibinfo {volume} {88}},\
  \bibinfo {pages} {035209} (\bibinfo {year} {2013})},\ \Eprint
  {https://arxiv.org/abs/1305.4351} {arXiv:1305.4351 [nucl-th]} \BibitemShut
  {NoStop}%
\bibitem [{\citenamefont {Ronchen}\ \emph {et~al.}(2013)\citenamefont
  {Ronchen}, \citenamefont {Doring}, \citenamefont {Huang}, \citenamefont
  {Haberzettl}, \citenamefont {Haidenbauer}, \citenamefont {Hanhart},
  \citenamefont {Krewald}, \citenamefont {Meissner},\ and\ \citenamefont
  {Nakayama}}]{Ronchen:2012eg}%
  \BibitemOpen
  \bibfield  {author} {\bibinfo {author} {\bibfnamefont {D.}~\bibnamefont
  {Ronchen}}, \bibinfo {author} {\bibfnamefont {M.}~\bibnamefont {Doring}},
  \bibinfo {author} {\bibfnamefont {F.}~\bibnamefont {Huang}}, \bibinfo
  {author} {\bibfnamefont {H.}~\bibnamefont {Haberzettl}}, \bibinfo {author}
  {\bibfnamefont {J.}~\bibnamefont {Haidenbauer}}, \bibinfo {author}
  {\bibfnamefont {C.}~\bibnamefont {Hanhart}}, \bibinfo {author} {\bibfnamefont
  {S.}~\bibnamefont {Krewald}}, \bibinfo {author} {\bibfnamefont {U.~G.}\
  \bibnamefont {Meissner}},\ and\ \bibinfo {author} {\bibfnamefont
  {K.}~\bibnamefont {Nakayama}},\ }\href
  {https://doi.org/10.1140/epja/i2013-13044-5} {\bibfield  {journal} {\bibinfo
  {journal} {Eur. Phys. J.}\ }\textbf {\bibinfo {volume} {A49}},\ \bibinfo
  {pages} {44} (\bibinfo {year} {2013})},\ \Eprint
  {https://arxiv.org/abs/1211.6998} {arXiv:1211.6998 [nucl-th]} \BibitemShut
  {NoStop}%
\bibitem [{\citenamefont {Skawronski}\ \emph {et~al.}(2019)\citenamefont
  {Skawronski}, \citenamefont {Blankleider},\ and\ \citenamefont
  {Kvinikhidze}}]{Skawronski:2018yhu}%
  \BibitemOpen
  \bibfield  {author} {\bibinfo {author} {\bibfnamefont {T.}~\bibnamefont
  {Skawronski}}, \bibinfo {author} {\bibfnamefont {B.}~\bibnamefont
  {Blankleider}},\ and\ \bibinfo {author} {\bibfnamefont {A.}~\bibnamefont
  {Kvinikhidze}},\ }\href {https://doi.org/10.1103/PhysRevC.99.034001}
  {\bibfield  {journal} {\bibinfo  {journal} {Phys. Rev. C}\ }\textbf {\bibinfo
  {volume} {99}},\ \bibinfo {pages} {034001} (\bibinfo {year} {2019})},\
  \Eprint {https://arxiv.org/abs/1810.06035} {arXiv:1810.06035 [nucl-th]}
  \BibitemShut {NoStop}%
\bibitem [{\citenamefont {Meissner}(1995)}]{Meissner:1995ra}%
  \BibitemOpen
  \bibfield  {author} {\bibinfo {author} {\bibfnamefont {T.}~\bibnamefont
  {Meissner}},\ }\href {https://doi.org/10.1103/PhysRevC.52.3386} {\bibfield
  {journal} {\bibinfo  {journal} {Phys. Rev. C}\ }\textbf {\bibinfo {volume}
  {52}},\ \bibinfo {pages} {3386} (\bibinfo {year} {1995})},\ \Eprint
  {https://arxiv.org/abs/nucl-th/9506030} {arXiv:nucl-th/9506030 [nucl-th]}
  \BibitemShut {NoStop}%
\end{thebibliography}%

\end{document}